\documentclass[preprint]{aastex} 
\usepackage{graphicx}

\newcommand{\be}{\begin{equation}}
\newcommand{\ee}{\end{equation}}

\newcommand{\taudamp}{ { \tau_{\rm damp}}} 
\newcommand{\taudisk}{ { \tau_{\rm disk}}} 
\newcommand{\taued}{ { \tau_{\rm ed}}} 
\newcommand{\taucirc}{ { \tau_{\rm circ}}} 
\newcommand{\vej}{ { v_{\rm ej}}} 
\newcommand{\siniobs}{{ \sin i_{obs} }} 

\begin{document}

\title{GIANT PLANET MIGRATION THROUGH THE ACTION OF \\ 
DISK TORQUES AND PLANET-PLANET SCATTERING} 

\medskip
\author{Althea V. Moorhead$^1$ and Fred C. Adams$^{1,2}$} 
\bigskip 

\affil{$^1$Michigan Center for Theoretical Physics \\ 
Physics Department, University of Michigan, Ann Arbor, MI 48109}

\affil{$^2$Astronomy Department, University of Michigan, Ann Arbor, MI 48109}

\begin{abstract} 

This paper presents a parametric study of giant planet migration
through the combined action of disk torques and planet-planet
scattering. The torques exerted on planets during Type II migration in
circumstellar disks readily decrease the semi-major axes $a$, whereas
scattering between planets increases the orbital eccentricities
$\epsilon$. This paper presents a parametric exploration of the
possible parameter space for this migration scenario using two
(initial) planetary mass distributions and a range of values for the
time scale of eccentricity damping (due to the disk). For each class
of systems, many realizations of the simulations are performed in
order to determine the distributions of the resulting orbital elements
of the surviving planets; this paper presents the results of
$\sim8500$ numerical experiments. Our goal is to study the physics of
this particular migration mechanism and to test it against
observations of extrasolar planets.  The action of disk torques and
planet-planet scattering results in a distribution of final orbital
elements that fills the $a-\epsilon$ plane, in rough agreement with
the orbital elements of observed extrasolar planets. In addition to
specifying the orbital elements, we characterize this migration
mechanism by finding the percentages of ejected and accreted planets,
the number of collisions, the dependence of outcomes on planetary
masses, the time spent in 2:1 and 3:1 resonances, and the effects of
the planetary IMF.  We also determine the distribution of inclination
angles of surviving planets and the distribution of ejection speeds
for exiled planets.

\end{abstract} 

\medskip 
$\,$  


\bigskip 

\section{Introduction} 

With over one hundred extrasolar planets discovered thus far, the past
decade has witnessed a revolution in our understanding of planetary
systems. The initial discoveries (Mayor and Queloz 1995; Marcy and
Butler 1996) showed that the orbital elements of the extrasolar
planets are significantly different from those of the planets in our
solar system.  Some fraction of the giant planets are found in short
period orbits with $P_{\rm orb} \approx 4$ days (semi-major axes $a
\approx$ 0.05 AU), while others display longer period orbits of high
eccentricity 0 $\le \epsilon \le$ 0.9 (where the high end of this
range remains somewhat uncertain). Subsequent discoveries have shown
that such planetary systems are relatively common and have a rich
variety of architectures (e.g., Marcy and Butler 1998, 2000; Hatzes et
al.  2000; Perryman 2000). In particular, the observed planets
populate a large portion of the $a-\epsilon$ plane. An important
astronomical challenge is to provide a theoretical explanation for the
observed distributions of orbital elements. A related challenge is to
understand the physical mechanism through which planets migrate inward
from their birth sites.

This paper presents a theoretical study of giant planet migration
through the combined action of disk torques and scattering by other
planets in the system. We consider planets of Jovian mass (in order of
magnitude) so that the planets clear gaps in the disk and thus
experience Type II migration. During the epoch of planet formation and
migration, both gaseous circumstellar disks and multiple planets are
expected to be present.  Sufficiently massive disks are effective at
exerting torques on planets and moving them inward, thereby changing
their semi-major axes $a$. Scattering interactions between planets are
effective at increasing the orbital eccentricities $\epsilon$ (this
point was emphasized in Adams and Laughlin 2003; hereafter AL2003).
This paper presents a statistically comprehensive study of this
migration mechanism and demonstrates that the interplay between these
two effects leads to a rich variety of possible outcomes. Because
these systems cover a wide range of parameter space and tend to be
chaotic, this process results in a broad distribution for the orbital
elements of the final systems.  This model -- Type II migration driven
by interactions with a circumstellar disk and by dynamical scattering
from other planets -- naturally produces the entire possible range of
semi-major axis $a$ and eccentricity $\epsilon$.

This research builds on an extensive body of previous work. Migration
was anticipated long before extrasolar planets were detected (e.g.,
Goldreich and Tremaine 1980; Lin and Papaloizou 1993) and a host of
different migration mechanisms have been developed (see the review of
Ward and Hahn 2000). Type I migration occurs when a planet (or forming
planet) is too small to clear a gap in the disk and remains embedded
within the fluid (Ward 1997); the planet drives wakes into the
background gas of the disk and these wakes, in turn, exert torques on
the planet and push it inward. In this study, we assume that the
planets are already formed and are large enough to clear gaps in the
disk, so that the starting point of these calculations takes place
after Type I migration has run its course (although it remains
possible for these early stages to provide an alternate explanation of
the observed orbital elements).  This paper considers a parametric
treatment that corresponds to Type II migration, wherein the planet
clears a gap in the gaseous disk. Many studies of Type II migration
have been carried out to explain the newly discovered extrasolar
planets in short period orbits (e.g., Lin et al. 1996; Trilling et
al. 1998; Nelson et al. 2000; Bryden et al. 2000; Bate et al. 2003),
although such models generally do not readily explain the large
eccentricities observed in many planetary orbits. For completeness we
note that additional models of runaway migration have been proposed as
a way to explain ``hot Jupiters'' (sometimes called Type III migration
-- see Masset and Snellgrove 2001; Masset and Papaloizou 2003),
although they are not considered here.

Another way to achieve shorter periods is through gravitational
scattering interactions with a disk of planetesimals (Murray et
al. 1998), although this mechanism requires a great deal of mass in
solid materials inside the orbit of the giant planet (see also the
Appendix to AL2003). A complementary collection of papers has studied
migration through planet-planet scattering alone, starting with two
planets (Rasio and Ford 1996; Weidenschilling and Marzari 1996; Katz
1997) and continuing with more complicated configurations (Ford et
al. 2001; Marzari and Weidenschilling 2002; Papaloizou and Terquem
2002; AL2003). These studies indicate that planet scattering is highly
effective at producing large orbital eccentricities but does not
readily move planets inward to small semi-major axes.  Specifically,
the ratio of the initial to final semi-major axes, $a_0/a_f$, is
typically only 3 -- 5, and a larger ratio would require a great deal
of additional mass in scattering bodies (more than is thought to be
available in such disks).

In light of the aforementioned research, one economical way to explain
both the observed (small) semi-major axes and the observed (large)
eccentricities is through a combined model with disk torques and
multiple planets (Lin and Ida 1997; Kley 2000; AL2003; Thommes and
Lissauer 2003; Kley et al. 2004). Many of the previous studies focus
on explaining particular observed two-planet systems like GJ876 (e.g.,
Snellgrove et al. 2001; Lee and Peale 2002; Murray et al. 2002) and 47
UMa (Laughlin et al. 2002). This paper adopts a more general treatment. 

This present work has two modest goals: The first objective is to
explore the physics of this migration mechanism by extending previous
calculations to encompass a wider range of parameter space.  This work
is a straightforward generalization of AL2003, but extends that paper
in several ways: [1] In addition to the random mass distribution of
AL2003, this paper considers a a log-random initial mass function for
the planets. [2] This paper explores a much wider range of time scales
for eccentricity damping due to the disk. [3] This paper includes
starting configurations that lead to the planets being initially
caught in both the 2:1 and 3:1 mean motion resonances, and we track
how long the planets stay near resonance.  [4] The distributions of
ejection velocities for escaping planets are determined. [5] In order
to isolate the effects of the input parameters on the final results,
this paper presents the orbital elements both immediately after
planets are lost and after corrections for additional evolution are
taken into account.  [6] This work includes a tenfold increase in the
number of numerical simulations and hence in coverage of parameter
space (for a total of $\sim8500$ trials).

The second goal of this paper is to determine if this migration
mechanism can account for the orbital elements of the observed
extrasolar planets. Since the observed orbital elements of these
planetary systems explore (nearly) the full range of possible
semi-major axis and eccentricity, filling the $a-\epsilon$ plane is a
necessary, but not sufficient, condition on a complete theory of
planet migration. The mechanism explored here can be made consistent
with the observed orbital element distributions, but such a comparison
is preliminary and caution should be taken.

\section{Methods and Initial Conditions} 

This section outlines our basic migration model which combines the
dynamical interactions between two planets with inward forcing driven
by tidal interactions with a background nebular disk, i.e., Type II
migration (see also Kley 2000; Murray et al. 2002; Papaloizou 2003;
Kley et al. 2004).  Our goal here is to build on previous studies by
producing a statistical generalization of the generic migration
problem with two planets and an exterior disk -- a situation that we
expect is common during the planet formation process.

The numerical experiments are set up for two planets with the
following orbital properties: Two planets are assumed to form within a
circumstellar disk with initial orbits that are widely spaced. The
central star is assumed to be of solar-type with mass $M_\ast$ = 1.0
$M_\odot$.  For the sake of definiteness, the inner planet is always
started with orbital period $P_{in}$ = 1900 days, which corresponds to
a semi-major axis $a_{in}$ $\approx$ 3 AU.  This radial location lies
just outside the snowline for most models of circumstellar disks and
thus provides a fiducial starting point where the innermost giant
planets are likely to form.  For most of the simulations, the second
(outer) planet is placed on an orbit with the larger period $P_{out} =
\pi 2^{1/4} P_{in} \approx (3.736\dots) P_{in}$. With this starting
state, the planets are not initially in resonance but will first
encounter the 3:1 resonance as the outer planet migrates inward.  As
the system evolves, the two orbits become closer together. With these
starting states, the planets are sometimes caught in the 3:1
resonance, but often pass through and approach the 2:1 resonance.  In
an alternate set of starting states, the outer planet is given an
initial orbital period $P_{out} = e P_{in} \approx (2.718\dots)
P_{in}$ so that the planets start inside the 3:1 resonance but outside
the 2:1 resonance.  In either case, the two planets are often caught
in mean motion resonances for some portion of their evolution (for a
more detailed description, see Lee and Peale 2002). In practice, the
initial period ratio is likely to have a distribution, but this paper
focuses on these two specific choices.  The initial eccentricities of
both planets are drawn from a uniform random distribution in the range
$0 < \epsilon < 0.05$. The planets are also started with a small, but
nonzero inclination angle in the range $i \le 0.03$ (in
radians). Planetary systems started in exactly the same orbital
plane tend to stay co-planar, whereas small departures such as these
allow the planets to explore the full three dimensions of space.

In this study we use two different distributions for the starting
planetary masses. We denote the planetary mass distribution as the IMF
(the initial mass function) where it should be understood that we mean
planet masses (not stellar masses). The first IMF is a uniform random
distribution in which the planet masses $m_P$ are drawn independently
from the range $0 < m_P < 5 m_J$, where $m_J$ is the mass of Jupiter.
In the second mass distribution, denoted as the log-random IMF, the
logarithm of the planet mass $\log_{10} [m_P/m_J]$ is drawn
independently from the interval $-1 \le \log_{10} [m_P/m_J] \le 1$.
The random mass distribution provides a good starting point to study
the physics of these systems -- it provides a good sampling of the
possible masses and mass ratios that two planet systems can have. On
the other hand, the observed distribution of planet masses is much
closer to a log-random distribution, so this latter distribution
provides a better model for comparison with observations. One result
of this work is a determination of how this migration mechanism
changes the planetary IMF, and it is useful to study this evolution
from the two different starting distributions.

The numerical integrations are carried out using a Bulirsch-Stoer
scheme (Press et al. 1986). The equations of motion are those of the
usual three body problem (two planets and the star) with the following
additional forcing terms: The circumstellar disk exerts torques on the
planets which lead to both orbital decay (Type II migration) and
damping of eccentricity. The star exerts tidal forces on the planets
which leads to additional energy dissipation and partial
circularization of the orbits.  Finally, the leading order curvature
of space-time (due to general relativity) is included to properly
account for the periastron advance of the orbits.

The outer planet in the system is tidally influenced by a background
circumstellar disk. Since the planets are (roughly) of Jovian mass,
they clear gaps in the disk and experience Type II migration. Instead
of modeling the interaction between the outer planet and disk in
detail, we adopt a parametric treatment that introduces a frictional
damping term into the dynamics. This damping force has the simple form
${\bf f}= -{\bf v} \taudamp^{-1}$ and is applied to the outer planet
at each time step, as a torque ${\bf r} \times {\bf f}$, so the outer
planet is gradually driven inward. The assumed damping force is
proportional to the velocity and defines a disk accretion time scale
$\taudamp$. We assume here that the disk inside the orbit of the outer
planet is sufficiently cleared out so that the inner planet does not
usually experience a Type II torque. Over most of its evolution, the
inner planet has a sufficiently small eccentricity so that it lies
well inside the (assumed) gap edge and receives negligible torque from
the disk (which lies outside the outer planet). When the inner planet
attains a high eccentricity, however, it can be found at a radius
comparable to that of the outer planet and can thus experience some
torque. This (relatively minor) effect is included by giving the inner
planet a torque that is reduced from that of the outer planet by a
factor $(r_{in}/r_{out})^6$.

In this set of simulations, we set the accretion time scale to be
$\taudamp$ = 0.3 Myr, consistent with recent estimates of this
quantity.  We can compare this time scale to several reference points:
[1] For example, Nelson et al. (2000) advocate migration time scales
of $10^4$ orbits for Jovian mass planets. [2] If disk accretion is
driven by viscous diffusion and can be described by an $\alpha$
prescription, then the disk accretion time scale $\taudisk =
\varpi^2/\nu$, where the viscosity $\nu = (2/3) \alpha v_T^2
\Omega^{-1}$ (Shu 1992). The disk scale height $H$ can be written in
the form $H$ = $v_T/\Omega$, where $v_T$ is the sound speed, and the
accretion time becomes $\taudisk = 1.5 (\varpi/H)^2 \Omega^{-1}
\alpha^{-1}$. If we evaluate the disk scale height $H$ and rotation
rate $\Omega$ for a temperature of $T = 70$ K at $\varpi$ = 7 AU
(where the outer planet forms and begins its migration), the adopted
disk accretion time scale $\taudamp$ = 0.3 Myr corresponds to $\alpha$
= $7 \times 10^{-4}$. This value falls comfortably within the expected
range $10^{-4} \le \alpha \le 10^{-2}$ (see Shu 1992). [3] As another
point of comparison, three-dimensional simulations of Jovian planets
in circumstellar disks (Kley, D'Angelo, and Henning 2001) find similar
migration time scales, about 0.1 Myr, which agree with two-dimensional
simulations done previously (Kley 1999). In these numerical studies,
the disks have slightly larger $\alpha$ = $4 \times 10^{-3}$ (hence
the slightly shorter time scale), scale height $H/r$ = 0.05, and disk
mass $M_d$ = $3.5 \times 10^{-3} M_\odot$ within the annulus 2 AU $\le
r \le$ 13 AU.  Note that the disk mass must be larger than the planet
masses in order to drive Type II migration. Notice also that the
migration time scale is assumed to be independent of the orbital
eccentricity, although more complicated behavior is possible.

These simulations include an additional forcing term that damps the
eccentricity of the outer planetary orbit (as suggested by numerical
simulations of these systems). In other words, the same angular
momentum exchange between the disk and the planet that leads to
orbital migration can also modify the eccentricity of the orbit.
Unfortunately, previous work on this issue presents rather divergent
points of view. Most numerical studies indicate that the action of
disk torques leads to damping of eccentricity, and these results are
often supported by analytic calculations (e.g., Snellgrove et al.
2001; Sch{\"a}fer et al. 2004). On the other hand, competing analytic
calculations indicate that eccentricity can be excited through the
action of disk torques and this mechanism has been proposed as an
explanation for the observed high eccentricities in the extrasolar
planetary orbits (e.g., Ogilvie and Lubow 2003; Goldreich and Sari
2003; Papaloizou et al. 2001). One reason for this ambiguity is that
the interaction between the disk and the planet can be broken down
into the action of resonances in the disk, where the non-coorbital
corotation resonances act to damp the eccentricity of the planetary
orbits while the non-coorbital Lindblad resonances act to pump it
up. The net effect depends on a close competition between the damping
terms and the excitation terms.  In rough terms, the conditions that
result in eccentricity damping are those that lead to relatively
narrow gaps, which in turn correspond to large disk viscosity ($\alpha
\sim 10^{-3}$) and modest sized planet masses ($m_P \sim m_J$) as
assumed here. The disk surface density and scale height also play a
role (Bryden et al. 2000).  If the gap is not completely clear, then
the corotation resonance locations within the gap will contain gas
that can interact with the planet and help enforce eccentricity
damping (S.  Lubow, private communication; this issue is quite
complicated and warrants further discussion --- see Ogilvie and Lubow
2003, Goldreich and Sari 2003). In contrast, wide and clear gaps,
which result from smaller viscosity and/or larger planet masses ($m_P
\approx 10 - 20 m_J$), can lead to eccentricity excitation (Snellgrove
et al. 2001; Papaloizou et al. 2001).

In light of these ambiguities, we incorporate the effects of
eccentricity damping in a parametric manner. For completeness, we note
that the damping force described above (that which enforces inward
migration) also tends to damp the eccentricity, although this effect
is much smaller then the explicit eccentricity damping terms included
here.  Specifically, the orbital eccentricity of the outer planet is
damped on a time scale $\taued$, which is considered as a free
parameter in this treatment.  The eccentricity damping is enforced by
converting the cartesian variables to orbital elements
$(a,\epsilon,i)$, applying the damping term, and then converting back.
The inclination angle is not explicitly damped, although the outer
planet experiences a small damping effect due to the form of the
migration force. In this work, we explore a range of damping times
scales 0.1 Myr $\le \taued \le \infty$, where the $\taued \to \infty$
limit corresponds to no eccentricity damping.  We have also run test
cases in which $\taued$ varies with orbital eccentricity, so that more
eccentric orbits are damped to a greater extent, although the results
are not markedly different. Our numerical exploration of parameter
space suggests that the most relevant variable is the ratio of
eccentricity damping time to disk accretion time, where this ratio
falls in the range $1/3 \le \taued/\taudamp \le \infty$ for the
simulations presented here. For comparison, the full range of positive
values for this ratio considered in the literature is approximately
$0.01 \le \taued/\taudamp \le \infty$ (with an additional range of
negative values corresponding to eccentricity excitation). This paper
considers the more limited range because the behavior outside our
range is known: For small values of $\taued/\taudamp$, eccentricity
damping is highly efficient, few planets are ejected, and large
eccentricities are not produced (e.g., Lee and Peale 2002; Thommes and
Lissauer 2003).  For negative values of $\taued/\taudamp$,
eccentricity is excited. We find that even with no eccentricity
damping, this model tends to overproduce eccentricity relative to the
currently observed sample of extrasolar planets; eccentricity
excitation could lead to even larger discrepancies. Note that one
advantage of this parametric treatment is that thousands of
simulations can be performed and the full distributions of final
orbital elements can be determined.

The numerical code includes relativistic corrections to the force
equations (e.g., Weinberg 1972). This force contribution drives the
periastron of both planetary orbits to precess (in the forward
direction).  Because the effect is greater close to the star, the
inner planet experiences more precession, and the net effect is to
move the two planets away from resonance. If the planets migrate
sufficiently close to the central star, this differential precession
effect can keep the planets out of a perfect resonance. Since resonant
conditions lead to greater excitation of orbital eccentricity, which
in turns drives the system toward instability, this relativistic
precession acts to make planetary systems more stable. In these
simulations, however, the planets rarely migrate close enough to the
star to make this effect important, but it is included for
completeness.

The simulations also include energy lost due to tidal interactions
between the planets and their central stars. In these simulations, the
planets spend most of their time relatively far from the star where
tidal interactions are negligible. As a result, we adopt a simplified
treatment of this effect. Specifically, the force exerted on the
planet due to tidal interactions is written in the approximate form 
\be 
{\bf F} = - {G m_P R_\ast^5 \over C j r^{11} } \bigl[ 
r^2 {\bf v} - ({\bf r} \cdot {\bf v}) {\bf r} \bigr] 
{0.6 r_p^3 \over 1 + (r_p / R_\ast)^3 } \, , 
\ee
where $R_\ast$ is the radius of the star, $r_p$ is the distance of
closest approach for a parabolic orbit with angular momentum $j$, and
$C = 2 \sqrt{\pi}/3$ is a dimensionless constant of order unity (for
further discussion, see Papaloizou and Terquem 2001; Press and
Teukolsky 1977). This formula implicitly assumes that the time between
encounters is long compared to the time for tidal interaction itself
and that most of the forcing occurs near the point of closest approach. 
This approximation is valid when the close encounters occur due to
planetary orbits with high eccentricities, which is generally the case
for planets in these simulations. Note that for longer term evolution 
of close planetary orbits, such as circularization over Gyr time
scales, an alternate approximation for the tidal forces is necessary
(see Section 4.2).

The simulations allow for collisions to take place between the
planets, and between the planets and the star. The effective radius
for planetary collisions is taken to be $R_P = 2 R_J$, with cross
section $\sigma_P$ = $4 \pi R_J^2$, which implicitly assumes that the
planets have not fully contracted. In order to model accretion events,
we assume that when a planet wanders within a distance $d = 2 \times
10^{11}$cm of the central star, accretion takes place. This distance
corresponds to $d \sim 3 R_\odot$, which is a typical radius for solar
type stars during the early part of their pre-main-sequence phase of
evolution.

For a given set of starting conditions (described above), each
numerical experiment is integrated forward in time and the system
follows the same basic evolutionary trend (see Fig. \ref{fig:splot}
and Section 3): The planets are started with a sufficient separation
so that they have weak initial interactions and are far from
resonance.  As the outer planet migrates inward through the action of
disk torques, the planets often enter into a mean motion resonance,
usually the 3:1 or 2:1 resonance because of the starting conditions. 
The tendency to enter 3:1 versus 2:1 resonances varies with the
planetary IMF, with a linear IMF producing more planets in 3:1
resonances and a log IMF producing more 2:1 systems.  In addition, the
2:1 resonances last longer, implying that they are more stable. The
two planets then migrate inwards together, staying relatively close to
resonance, but displaying ever larger librations as the orbital
eccentricities of both planets increase (on average). The
eccentricities increase until the system (often) becomes unstable, and
a wide range of final system properties can result. In practice, we
continue the simulations until one of the following stopping criteria
is met: A planet is ejected, the planets collide with each other, a
planet is accreted by the central star, or a maximum integration time
limit is reached (set here to be 1.0 Myr). This latter time scale
represents the time over which the disk contains enough mass to drive
inward migration of planets; the disk could retain enough gas to
exhibit observational signatures over a longer time.

After a planet is lost (through ejection, accretion, or collision),
the numerical integration is stopped and the orbital elements of the
surviving planet are recorded. In general, however, the orbital
elements of the surviving planet can continue to evolve after a planet
is lost as long as the disk is still present. In order to separate the
effects of the combined migration mechanism (i.e., Type II migration
with planet scattering) from the additional evolution, we first
present the results with no additional evolution in the following
section. In order to compare with the observed orbital elements of
extrasolar planets, we consider possible algorithms for this
additional evolution in Section 4.

\section{Results from the Numerical Simulations} 

This paper presents the results of an ensemble of $\sim8500$
simulations that follow the early evolution of two-planet solar
systems subjected to disk torques using the methodology described
above. The simulations use two different planetary IMFs and, for each
IMF, four choices of the eccentricity damping time scale $\taued$. For
each set of these input parameters, we completed approximately 800 --
1000 solar system simulations. We then determined the resulting
distributions of semi-major axis $a$, eccentricity $\epsilon$,
inclination angle $i$, and surviving planetary mass $m_P$. These
results can be used to quantify the outcome of this migration
mechanism (see below) and can be compared to observed distributions of
orbital elements in extrasolar planetary systems (Section 4).

\subsection{Evolution of Orbital Elements}

\begin{figure}  
{\hskip -0.5truein {\epsscale{1.10} \plotone{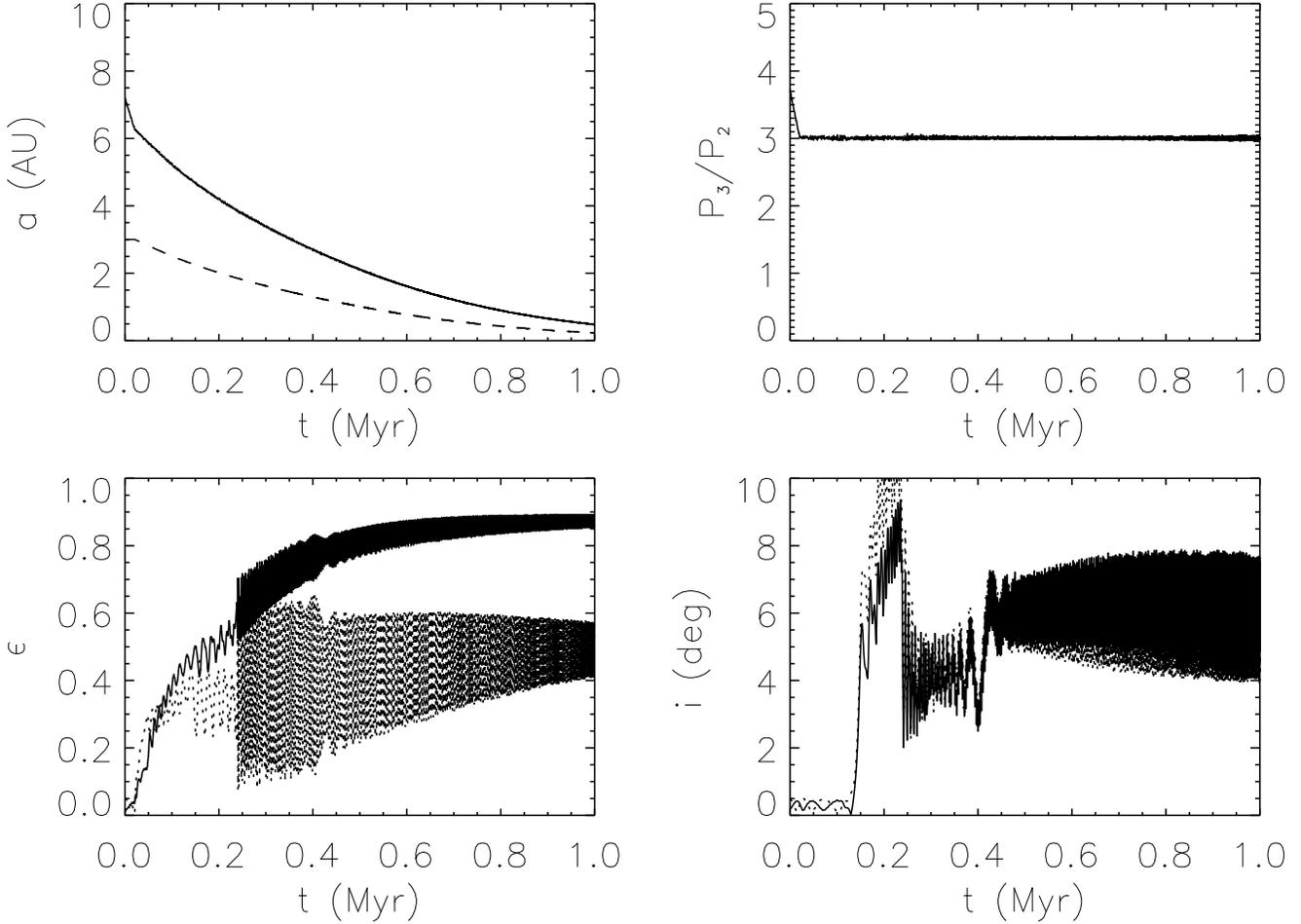} } }  
\caption{Time evolution of a typical system of two interacting
planets migrating under the influence of torques from a circumstellar
disk. The upper left panel shows the time evolution of the semi-major
axes, which decrease steadily on the migration time scale $\taudamp$. 
The upper right panel shows the ratio of the orbital periods. This
ratio quickly decreases to 3 and stays close to this value for much of
the evolution (the two planets are near the 3:1 resonance -- see
section 3.1). The evolution of eccentricity is illustrated in the
lower left panel, which shows that the eccentricity of both planets
steadily increases at first and then enters into a complicated time
series including both short period oscillations and an overall growth 
trend on longer time scales. The lower right panel shows the 
corresponding time evolution of the inclination angle. Both planets 
wander back and forth out of the original orbital plane, but the 
inclination angles vary by only a few degrees. }
\label{fig:splot} \end{figure}

To illustrate the general behavioral trend of these systems, we follow
the evolution of orbital elements for a collection of representative
simulations. The result of one such run is shown in Fig. \ref{fig:splot}.
The first panel shows how the semi-major axis of each planet decreases
smoothly with time; this basic trends holds for essentially all cases.
In the second panel of Fig. \ref{fig:splot}, we plot the period ratio
of the two planets, and find that it quickly approaches and remains
near 3.  This result indicates that the two planets may be in a 3:1
mean motion resonance (see Section 3.3 for further discussion).  This
behavior occurs during the early evolution for the majority of cases,
although in some cases the outer planet passes through the 3:1 period
ratio (and hence the 3:1 resonance) and the period ratio remains near
$\sim$2 for most of the evolution. In other systems, the period ratio
remains near 3 for the early evolution, and then the planets move
through the 3:1 resonance, become closer, and reside near the 2:1
resonance for the latter part of the simulation.  A more detailed
accounting of how long various systems spend near the 3:1 and 2:1
resonances is given below (Section 3.3).

The behavior of eccentricity and inclination angle is more complex.
We find that orbital eccentricity increases steeply in the beginning,
and then undergoes oscillations about an average value that increases
at a slower rate. The eccentricity exhibits varying behaviors over
different spans of time and varies substantially from case to case.
In the more stable systems, the eccentricity values level off and
experience variations about the mean. In unstable cases, the
eccentricities are driven to ever larger values until a planet is
either ejected or accreted onto the central star. The inclination
angle experiences a similar evolutionary trend.

\begin{figure}  
{\hskip -0.5truein {\epsscale{1.10} \plotone{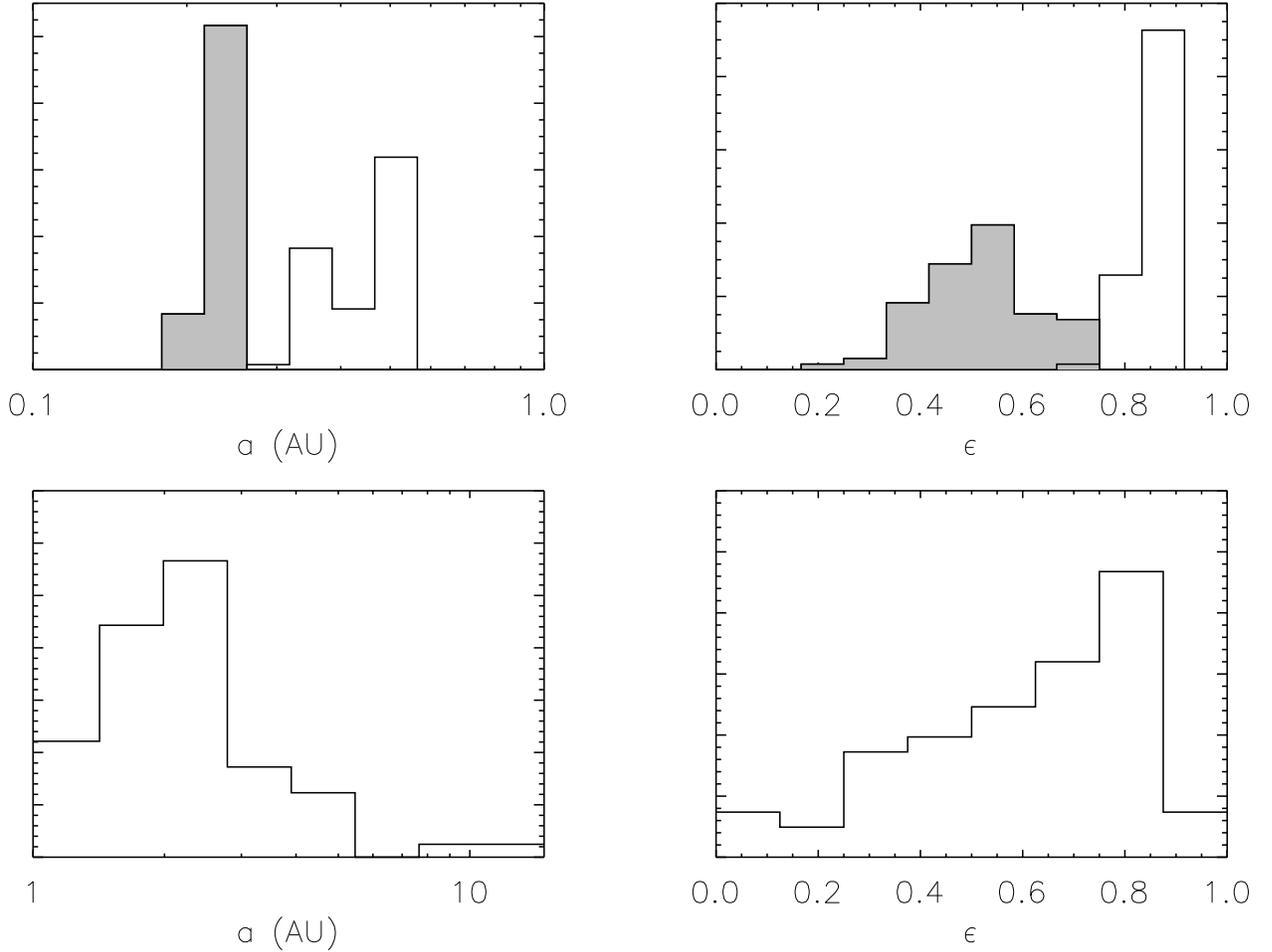} } }  
\caption{Illustration of sensitive dependence on initial conditions.
The top panels show the results of simulations performed for two equal
mass planets with $m_P = m_J$. The set of simulations use the same
starting conditions except for the choice of angular orbital elements.
In all cases, both planets survive to the end of the fiducial time
period of 1 Myr, but the orbital elements of the planets are
different, i.e., they show a distribution of values. The orbital
elements of the inner planet are shown as the shaded histogram; those
of the outer planet correspond to the unshaded histrogram.  The bottom
two panels show analogous results for simulations done with two equal
mass planets with $m_P = 5 m_J$.  In this case, one of the planets is
always ejected, but the remaining planet takes on a distribution of
values for its orbital elements. } 
\label{fig:sensdep} 
\end{figure}

This complex behavior of the orbital element leads to a distribution
of final values. To illustrate this sensitive dependence on the
initial conditions, we have run a set of numerical experiments with
equal mass planets, an eccentricity damping time scale $\taued$ = 1
Myr, and starting configurations with the outer planet outside the 3:1
resonance. The starting values of the angular orbital elements are
chosen to vary randomly from case to case.  The results are shown in
Fig. \ref{fig:sensdep} for the case of two $m_P = 1.0 m_J$ planets
(top two panels) and two $m_P = 5.0 m_J$ planets (bottom two panels).
For the case of smaller planets (1 $m_J$), both planets survive the
entire 1.0 Myr time span of the simulations, but they attain a wide
distribution of final orbital elements. For the case of larger planets
($5 m_J$), one of the planets is always ejected, whereas the remaining
planet attains a distribution of orbital elements as shown.
Fig. \ref{fig:sensdep} shows that effectively equivalent starting
conditions lead to a well-defined distribution of outcomes. In other
words, for this class of simulations, the outcomes must be described
in terms of the distributions of $a$ or $\epsilon$, rather than as 
particular values of $a$ or $\epsilon$.

\subsection{End State Probabilities} 

The simulations end in a variety of different states, including
ejection, accretion, collision, or the survival of both planets past
the fiducial time span of one million years. The frequencies of each
fate are listed in Table 1 for varying eccentricity damping time
scales and for both planetary IMFs. The number of ejected planets
depends sensitively on the IMF: Only one third of the planets were
ejected for a logarithmic IMF, whereas more than half were ejected for
a linear IMF. This behavior is expected because the linear IMF
provides more massive planets, which in turn produce a disturbing
function of greater magnitude and lead to more frequent ejection. We
also find that the outer planet is more than 3--4 times as likely to
be the ejected planet, and the inner planet is almost always the
accreted planet. The incidence of each end state exhibits no clear
trend with respect to eccentricity damping time scale (for a given
planetary IMF). For the case in which the planets are started just
outside the 2:1 resonance (inside the 3:1 resonance), the end state
probabilities are similar, but show a slight preference for accretion
relative to ejection (see Table 1).

Averaged over all outcomes, the mean time of the simulations (which
end when a planet is lost) is about 0.5 Myr; this time scale is
roughly comparable to the viscous damping time of $\taudamp$ = 0.3
Myr. Accretion events take the longest, with an average time of 0.55
Myr; ejection events have a mean time of 0.22 Myr; collisions take
place the fastest with a mean time of only 0.10 Myr. 

The end states depend on the planet masses in a systematic way, as
shown in Figs. \ref{fig:massplot} and \ref{fig:massplotalt}, which
show the various possible end states as a function of the masses.  For
systems in which the outer planet is substantially more massive than
the inner one, $m_{out} \gg m_{in}$, the inner planet is nearly always
driven to high eccentricities and accreted onto the central star (as
shown by the blue diamonds in Figs. \ref{fig:massplot} and
\ref{fig:massplotalt}).  In the opposite limit where $m_{out} \ll
m_{in}$, the outer planet tends to be ejected (as shown by the green
squares in the figures).  When both planets are massive, corresponding
to the upper right portion of the mass plane, either planet can be
readily ejected. When both planets have relatively low masses,
corresponding to the lower left portion of the mass plane, both
planets tend to survive (shown by the cross symbols in the figures).

Tables 2 -- 4 present the distributions of mass $m_P$, semi-major axis
$a$, and eccentricity $\epsilon$ at the end of the simulations. Each
entry lists the mean value of the distribution as well as its width
(variance). Table 2 presents the planetary masses for all cases,
including planets that are lost (ejected planets and accreted
planets).  The following tables list the parameters that characterize
the distributions of semi-major axis (Table 3) and orbital
eccentricity (Table 4) for the planets that remain at the end of the
simulations.  For accretion or ejection events (of either planet), the
distributions of semi-major axis, eccentricity, and mass are roughly
similar for a given planetary IMF and varying $\taued$ (although
variations do exist, especially at the extremes of our chosen range of
$\taued$).  The collisions result in significantly different orbital
properties (not listed in the Tables), with smaller eccentricity
$\epsilon$ and larger mass $m_P$. The other general trend that emerges
from this suite of simulations is that the systems that remain stable
over the entire 1 Myr integration time are those with the smallest
planets, with a mean mass of only 0.79 $m_J$ (for the log-random IMF,
averaged over the four values of $\taued$) compared to a mean mass
$m_P$ = 2.8 $m_J$ for the whole ensemble.

For the log-random IMF, which is closest to producing the observed
mass distribution, roughly one third of the systems retain both
planets at the end of the 1 Myr integration time (see Table 1). For
comparison, about 10 -- 20 percent of the observed extrasolar
planetary systems show multiple planets (to date -- see
www.exoplanets.org).  However, the theoretical systems that retain
multiple planets tend to have planetary masses that are smaller than
average (Table 2) and such low mass planets may have escaped
detection. In addition, as many as half of the systems observed with
one planet may contain additional bodies (Fischer et al. 2001). More
data is necessary to determine whether or not this issue is
problematic for the theory.

\subsection{Behavior of Resonance Angles}

\begin{figure} 
{\hskip -0.5truein {\epsscale{1.10} \plotone{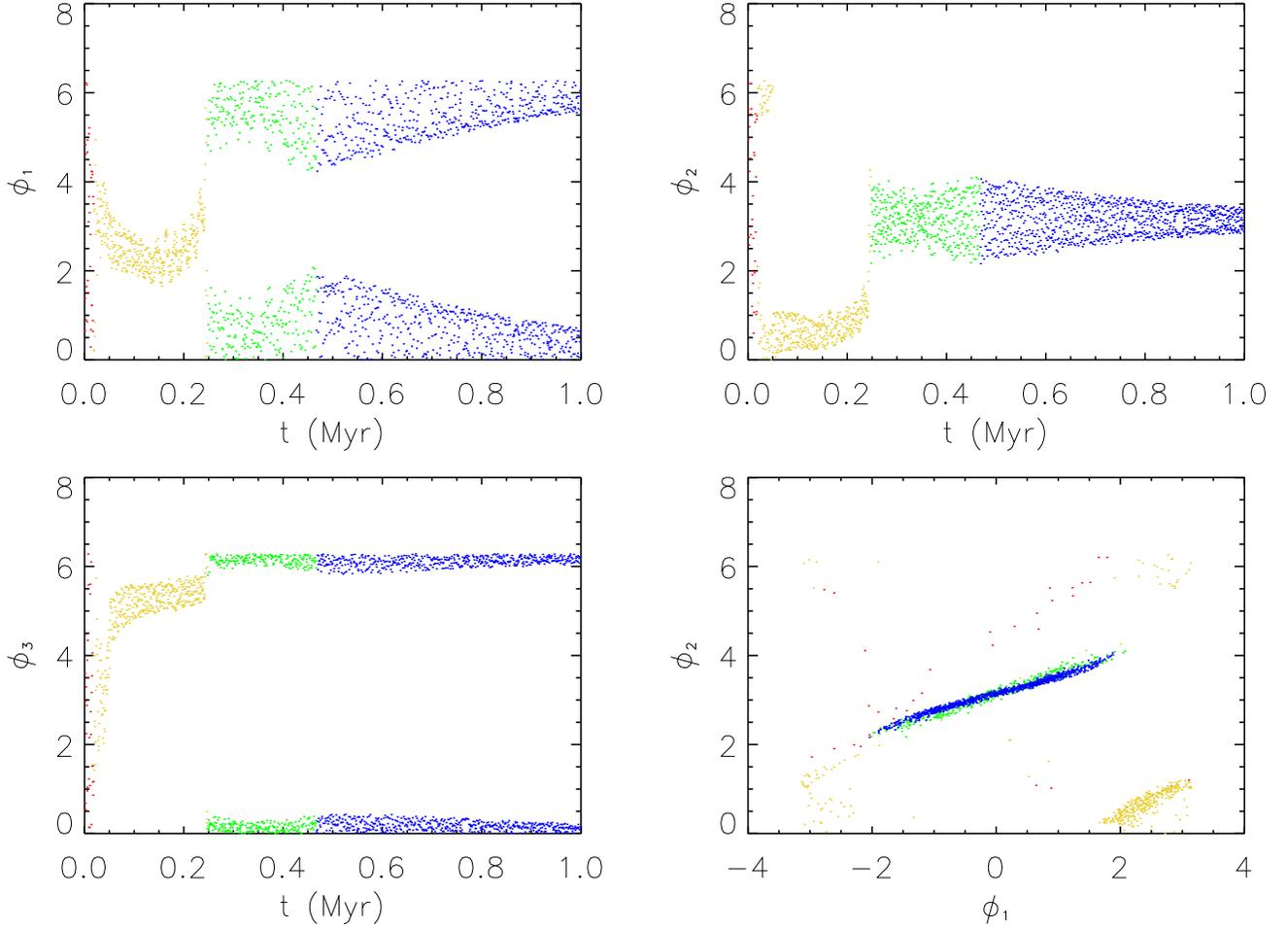} } }  
\caption{Representative behavior of the resonance angles. The first
three panels show the time evolution of the 3:1 resonance angles for a
representative simulation in which two equal mass planets with $m_P$ =
1.0 $m_J$ migrate inward together and approach the 3:1 mean motion
resonance. As shown here, the resonance angles exhibit complex
behavior and exhibit large librations about the resonance;
nonetheless, a well-defined resonant condition is reached.  The lower
right panel shows two of the resonance angles plotted against each
other. For most of the evolution, $t > 0.25$ Myr, the system librates
around the point $\phi_1$ = 0, $\phi_2 = \pi$. }
\label{fig:resonance} 
\end{figure}

As shown above, the period ratio of the two planets quickly and
smoothly approaches an integer value, often 3, and thus approaches a
mean motion resonance. As one benchmark, 70\% of the systems studied
here spend at least 10,000 years near the 3:1 resonance.  This result is
supported by the behavior of the 3:1 resonance angles over time. 
To illustrate this behavior we focus on the three angles
\be 
\phi_1 = 3 \lambda_2 - \lambda_1 - 2 \varpi_1 , \qquad 
\phi_2 = 3 \lambda_2 - \lambda_1 - \varpi_2 - \varpi_1 , \qquad 
\phi_3 = 3 \lambda_2 - \lambda_1 - 2 \varpi_2 , 
\ee
where the $\lambda_j$ are the mean longitudes and the $\varpi_j$ are
the longitudes of pericenter (Murray and Dermott 2001). Note that a 
complete description of the 3:1 resonance requires three additional
angles, although their behavior is similar to those considered here.
For the representative case of two planets of one Jupiter mass each,
Fig. \ref{fig:resonance} shows how the system passes through different
versions of the 3:1 resonances.  For example, between 50 and 200
thousand years, the first three angles librate about values that are
120 degrees apart.  Then, between 0.25 and 1 Myr, the first and third
angles librate about an angular value that is 180 degrees different
from the second.

A wide range of behavior is displayed in the time evolution of the
resonance angles, although the overall defining trend can be described
as follows (see also Beaug{\'e} et al. 2003; Ferraz-Mello et al. 2003;
Lee 2004): The planets tend to approach a mean motion resonance, but
generally exhibit large librations about a perfect resonant condition.
Two effects contribute to this behavior. The circumstellar disk exerts
a torque on the outer planet and acts to move the planet inward and
away from resonance; although the inner planet experiences a greatly
reduced torque, it is not enough to keep up, and the two planets must
continually readjust their orbital elements to maintain a resonant
condition. In addition, the planets are massive enough and
sufficiently close together so that they tend to excite the orbital
eccentricities; this continual pumping up of the eccentricities can
eventually act to make the system unstable.  Notice that when the
planets have low masses, they tend to stay in resonance longer.
Indeed, at the end of the simulations, the subset of solar systems
that retain both planets over the entire 1 Myr time period have a much
lower average mass (see Table 2, Figs. \ref{fig:massplot} and
\ref{fig:massplotalt}, and the previous subsection).

Although most of these two-planet systems spend some of their
evolutionary time in states with period ratio near 3:1, a substantial
fraction of the systems pass through the 3:1 resonance and approach a
2:1 period ratio. One should keep in mind that a rational period ratio
is a necessary, but not sufficient, condition for being in a mean
motion resonance. For the ensemble of simulations studied here, we
have kept track of the time for which the planets display period
ratios near 3:1 and near 2:1 (specifically, period ratios within $3
\pm 0.15$ and $2 \pm 0.15$, respectively). The results are compiled in
Table 5. The first four columns give the percentage of the time spent
in (near) each resonance, where the first value in each table entry
corresponds to the 2:1 resonance and the second value to the 3:1
resonance. Here, for the first four columns of results, the percentage
is calculated by integrating up the total time that any planet in the
given ensemble spends near the resonance and then dividing by the
total time that the ensemble of planets resides in the simulations.
Notice that this figure of merit is different from that obtained by
finding the percentage of time that each individual planet spends near
resonance, and then averaging that fraction over the planets; this
latter quantity is given in the last column in Table 5 for the total
samples (including all outcomes) for each planetary IMF.

For the linear IMF, Table 5 indicates that planets spend about 70
percent of their evolutionary time with period ratios of 3:1 and only
about 20 percent of their time with ratios near 2:1. The planets that
survive the longest (from the simulations that reach the stopping time
of 1 Myr without losing a planet) tend to reach the 2:1 resonance (see
below). As a result, the resonance fractions obtained by time
averaging over the whole ensemble of planets (the percentages given in
column 4 in Table 5) are more weighted toward the 2:1 resonance than
the fractions obtained by finding the individual resonance time
fractions and then averaging (the alternate percentages given in
column 5).  Similarly, for the log-random IMF, the systems spend an
average of 53 percent of their time with 2:1 period ratios and 38 
percent of their time with 3:1 period ratios (averaged over the four 
values of $\taued$). The main difference
between the two IMF choices is that the linear IMF has larger planets
and mass ratios closer to unity; our numerical results indicate that
this combination is more conducive to keeping the planets locked in
the 3:1 resonant condition. For those systems that stay near a period
ratio of 3:1 for more than 80 percent of their evolutionary time, the
distribution of mass ratios $m_{out}/m_{in}$ is sharply peaked near
unity, with a long tail to larger values. As a result, equal mass
planets tend to stay near resonance longer. Systems with with the 
inner planet more massive than the outer planet tend to move away from
resonance, often by ejecting the smaller planet, whereas systems with 
more massive outer planets often drive the smaller inner planet into 
the star. 

Another related result concerns the systems that survive for the
entire 1 Myr time span without ejecting a planet. As mentioned above,
the planets in these systems have relatively lower masses (Table 2).
For the log-random IMF, 69 percent of the surviving systems are found
with period ratios near 2:1, 28 percent of the systems show period
ratios near 3:1, and the remaining 3 percent are ``far'' from
resonance. Thus, surviving systems tend to be those that pass through
the 3:1 resonance and become locked into the 2:1 resonance. As
expected, the surviving systems show a mass distribution that is
weighted toward lower masses compared with the original log-random
distribution (which is roughly consistent with the mass distribution
of observed planets). The distribution of mass ratios $m_{out}/m_{in}$
is slanted toward values less than unity so that surviving systems
tend to have the outer planet less massive than the inner planet. For
the linear IMF, the results are somewhat different, where 67 percent
of the surviving systems have period ratios near 3:1 and the remaining
33 percent have 2:1 period ratios (see also Figs. \ref{fig:massplot} 
and \ref{fig:massplotalt}). 

\subsection{Distributions of Orbital Elements}

For each class of starting configuration, the simulations result in a
distribution of final system properties. The starting states can be
characterized by the planetary IMF, the eccentricity damping time
scale $\tau_{\rm ed}$, and the initial period of the outer planet
$P_{out}$ (although most of our simulations use $P_{out} \approx 3.736
P_{in}$). Each triple (IMF, $\tau_{\rm ed}$, $P_{out}$) thus leads to
distributions of final orbital elements, as reported in Tables 2 -- 4.
For each end state of the simulations (ejection, accretion, etc.),
these tables characterize the distributions of mass, eccentricity, and
semi-major axes, by specifying their mean values and widths. One must 
keep in mind the probabilistic nature of this type of problem. The 
simulations act as a mapping from an initial space to a final one, 
\be 
({\rm IMF}, \tau_{\rm ed}, P_{out}) \, \to \, 
\bigl\{ f_{out}, f_a (a), f_\epsilon (\epsilon), f_i (i), f_m (m_P), 
\dots \bigr\}  \, , 
\ee
where the entries on the left hand side are numbers (single values)
and the entries $f_j$ on the right hand side are distributions of
values (e.g., $f_{out}$ represents the fractional occurrence of each
outcome, $f_a(a)$ is the distribution of semi-major axes, etc.).

The resulting distributions of orbital elements, and planet masses are
plotted in Figs. \ref{fig:linhist} and \ref{fig:loghist}. For
comparison, these figures include the distributions of planet
properties from the observed sample (see Section 4).  The lower right
panel of each figure shows the mass distribution of the planets that
survive to the end of the simulations. The observed distribution of
planets provides us with $m_P \siniobs$ (rather than $m_P$); to
account for this ambiguity, we have used the quantity $2 m_P \siniobs$
as a working mass estimate for specifying the ``observed'' mass
distribution (notice also that the observational viewing angle is
different from the usual definition of inclination angle as an orbital
element so that $\siniobs \ne \sin i$). In spite of the tendency for
smaller planets to be ejected, the linear IMF tends to roughly
preserve its shape during the course of evolution (see Fig.
\ref{fig:linhist}).  However, in comparison, the mass distribution of
observed extrasolar planets has fewer high mass planets and more low
mass planets and is thus closer to the log-random distribution (which
we adopted as our second working IMF -- see Fig. \ref{fig:loghist}).
Even with the ejection of the smaller planets, the log-random IMF
model yields a final mass distribution that is close to the observed
distribution.  Some discrepancy occurs in the high mass tail, however,
because our log-random distribution has an upper bound at $m_P$ = 10
$m_J$. Our model also provides a small surplus of planets in the low
mass tail, relative to the observed mass distribution, although this
disagreement may be the result of less frequent detection of the
smallest planets.

The different sets of simulations (two planetary IMFs and four values
of the eccentricity damping time scale) tend to result in relatively
flat distributions of the semi-major axis (see the upper left panels
of Figs. \ref{fig:linhist} and \ref{fig:loghist}). For the case of no
eccentricity damping and a log-random IMF, however, the semi-major
axis distribution of the surviving planets displays a broad peak near
1 AU. As the eccentricity damping is increased (more damping with a
shorter time scale $\taued$), this peak becomes even broader (flattens
out) and moves toward lower values of $a$. Because Type II migration
torques are effective at moving planets inward, and because of the
random element introduced into the migration process (due to different 
starting angles and varying effective integration times), the
resulting values of semi-major axis fill the entire range covered by
current observations.

The distributions of eccentricities are shown in the upper right
panels in Figs. \ref{fig:linhist} and \ref{fig:loghist}. As
expected, the distribution of eccentricity shifts toward lower values
as the level of eccentricity damping is increased (i.e., as $\taued$
decreases). The general trend is for the simulations to excite the
orbital eccentricities to higher levels (averaged over the
distribution) than those found in the observational sample. The
exception to this rule occurs for the shortest eccentricity damping
time scale $\taued = 0.1$ Myr for the log-random planetary IMF: In
this class of systems, the resulting distribution of eccentricity is
shifted to lower values than the observed planetary orbits.  Taken at
face value (ignoring the possibility of selection effects in the
observational sample), this set of results argues that, within the
context of this migration scenario, the eccentricity damping time
scale cannot be smaller than about $\taued = 0.1$ Myr (otherwise,
resulting eccentricities would be too low) or larger than about
$\taued = 1.0$ Myr (otherwise, $\epsilon$ values would be too
large). We will return to this issue below.

The distributions of inclination angle are shown as the lower left
panels in Figs. \ref{fig:linhist} and \ref{fig:loghist}. The resulting
distributions of the inclination angle appear to be largely
independent of the eccentricity damping time scale $\taued$ for both
choices of planetary IMF. The distribution shows a well-defined peak
near $i \approx 6$ degrees for the random IMF and a broader peak near
$i = 3-5$ degrees for the log-random IMF. Although modest, these
angles are significantly larger than the starting inclination angles
($|i| \le 0.03 \approx 1.7$ degrees). For the shortest eccentricity
damping time scale and the log-random IMF (which has the greatest
number of small planets), however, the distribution of inclination
angle is shifted somewhat toward lower values.

As a general rule, increases of the inclination angle are correlated
with increases of eccentricity (consistent with the earlier studies of
AL2003; Thommes and Lissauer 2003).  For this ensemble of simulations,
we find that the inclination angle and the eccentricity have a linear
correlation coefficient in the range $r(N) \approx 0.33 - 0.66$ for
simulations that end in either ejection or accretion. Table 6 shows
these results for the eight classes of simulations conducted in this
study; also shown (in parentheses) are the numbers of simulations in
the sample used to compute each correlation coefficient. For the large
sample size ($N \sim 100 - 300$), these values of $r(N)$ are generally
considered significant (Press et al. 1986), but the correlation is not
exact. Notice that our simple treatment does not include the possible
damping of inclination angle by the circumstellar disk (e.g., Lubow
and Ogilvie 2001). Such damping would move the distributions of $i$ to
smaller values, but the correlations between eccentricity excitation
and inclination angle excitation would remain.

\subsection{Distribution of Ejection Speeds} 

A substantial fraction of the planetary systems eject a planet and
these planets can, in principle, be observed as free floating bodies.
Table 1 indicates that approximately one-third of the systems will
eject planets for the log-random planetary IMF, and over half of the
systems with a linear IMF will eject planets.  These planets have
roughly Jovian mass (Table 2) and thus do not represent a significant
mass contribution to the galaxy -- in other words, the number of
ejected planets is not large enough to be problematic.  However,
recent observations have found some evidence for freely floating
planets (e.g., Zapatero Osorio et al. 2000) and these predicted
planets must be consistent with the observations.

The planets are ejected with a well-defined distribution of speeds, as
shown in Fig. \ref{fig:veject}. For all cases considered here, the
distribution displays a well-defined peak near $\vej$ = 5 km/s and
most of the distribution falls between 0.5 and 20 km/s. For
comparison, the planets are ejected from orbits with semi-major axes
near 3 -- 7 AU, where the orbit speeds are about 11 -- 17 $\approx$ 14
km/s. The kinetic energy carried away by the ejected planets is thus a
small fraction ${\cal F}_E$ of the total. A rough estimate of this 
fraction is given by 
\be 
{\cal F}_E \approx {\vej^2/2 \over |E|} = {a \vej^2 \over GM_\ast } 
\approx (5/14)^2 \approx 0.13 \, , 
\ee 
where $\vej$ is the ejection speed and $a$ is the semi-major axis from
which ejection occurs.  This finding vindicates the assumption that
ejected planets tend to leave on (nearly) zero energy orbits (AL2003;
Marzari and Weidenschilling 2002).

We can understand the general form of the distribution of ejection
speeds with the following heuristic argument: Ejections occur through
close encounters between the planets. Let $b$ denote the impact
parameter of these interactions, so that the ejection speed can be
written 
\be 
{1 \over 2} \vej^2 = \alpha {G \langle m \rangle \over b} 
- {G M_\ast \over 2 a } \, , 
\ee
where $a$ is the semi-major axis of the ejected planet (before the
interaction), $\langle m \rangle$ is an average mass of the remaining
planet, and $\alpha$ is a dimensionless factor of order unity (which
depends on the geometry of the interaction). If we define a velocity
scale $v_0^2 \equiv G M_\ast /a$ and a length scale $r_0 \equiv 2 \alpha
(\langle m \rangle / M_\ast) a$, the ejection speed can be written in
the form 
\be
u = \bigl[ {1 \over \xi} - 1 \bigr]^{1/2} \, ,
\ee
where $u \equiv \vej/v_0$ and $\xi \equiv b/r_0$. If we assume that 
the impact parameter is distributed according to $dP \propto b db$ 
(the target area is circular), the probability distribution for the 
ejection speed takes the form 
\be 
{dP \over du} = {4 u \over (1 + u^2)^3 } \, . 
\label{eq:vdist} 
\ee 
As written, this probability distribution is normalized to unity over
the full range of dimensionless ejection speeds $0 \le u \le \infty$. 
In practice, the impact parameter has a minimum value given by the
radius of the (ejector) planet, $b_{\rm min}$ = $r_P$, and this value
implies a corresponding cutoff in the ejection speed $v_{\rm max}
\approx v_0 \sqrt{r_0/r_P}$.  However, the distribution falls rapidly
at high speeds so that the value of this cutoff is relatively 
unimportant. Equation (\ref{eq:vdist}) has the same form as the
distributions of ejection speeds found in the simulations (as shown in
Fig. \ref{fig:veject}). The model distribution has been normalized to 
agree with the simulations (note that the fraction of systems that 
experience ejection must be determined numerically). The distributions
agree if the velocity scale is taken to be $v_0 \approx 11$ km/s,
which implies that ejections (mostly) occur near the beginning of the
evolution (the outer planets, which are more often ejected, start near
7 AU where $(GM_\ast/a) \approx 11$ km/s).  Finally, we note that the
simple formula derived above assumes a single value for the velocity
scale. Since the planets can migrate inwards to different semi-major
axes before ejection, the true distribution will have a range of $v_0$
values; although this range is relatively narrow in the present
application, this effect tends to broaden the distribution of ejection
speeds relative to equation (\ref{eq:vdist}).

Under a reasonable set of assumptions, we can estimate the expected
population of free floating planets within a typical birth aggregate.
The velocity dispersion for a young star forming group/cluster is
about 1 km/s (Lada and Lada 2003; Porras et al. 2003). Given the
distribution of ejection speeds (Fig. \ref{fig:veject}), the majority
of ejected planets are predicted to be unbound to their birth
clusters.  As a first approximation, suppose that every solar system
produces migrating planets and that one third of them eject planets
(Table 1).  Of the ejected planets, about one tenth remain bound to
the group/cluster with ejection speeds $\vej < 1$ km/s. For every 30
stars in the cluster, it will thus contain one freely floating planet
that is gravitationally bound.  The unbound planets have ejection
speeds of $\sim$5 km/s. For an average cluster size of $R \sim 1$ pc,
the ejected planets would remain within their birth clusters for
$\sim$0.2 Myr.  If the young group/cluster remains intact for 10 Myr,
then 1/50th of the unbound planets will reside within the
group/cluster at any given time, and the cluster will contain one
unbound planet for every 150 stars. Given a fiducial group/cluster
size of $N_\ast \approx 300$ stars (Lada and Lada 2003; Porras et al.
2003), stellar birth aggregates will have $\sim10$ freely floating
planets at low speeds ($\vej \le 1$ km/s) and $\sim2$ freely floating
planets at higher speeds ($\vej \sim 5$ km/s) at any given time ($t
\le 10$ Myr).

\section{Comparison with Observed Extrasolar Planets} 

In order for a migration mechanism to be considered fully successful,
it must produce distributions of orbital elements that are consistent
with those of observed extrasolar planets. Given that the observed
distributions are incomplete and contain biases, however, and that
additional orbital evolution must take place between the end of our
simulations and the $\sim 1 - 6$ Gyr ages of the observed systems,
this type of comparison remains preliminary. In this section, we
briefly discuss the limitations of the data set and outline how the
distributions of orbital elements can evolve after the end of our
simulations. We then show that this migration mechanism meets the
necessary (but not sufficient) condition of being able to fill the
$a-\epsilon$ plane in a manner that is roughly consistent with
presently available data.

\subsection{Observed Sample of Extrasolar Planets} 

The observed sample of extrasolar planets used in this paper is taken
from the California and Carnegie Planet Search
Website.\footnote{http://exoplanets.org/science.html} In order to
compare theoretical results with this data set, some of its properties
must be taken into account. All planet searches using radial velocity
surveys are subject to selection effects. Since the surveys are
subject to a minimum (detectable) velocity amplitude, planetary
companions that induce reflex velocities that are too small cannot be
measured. This effect limits the sensitivity of the surveys to low
mass planets. In addition, planets with long periods (large semi-major
axes $a$) cannot be adequately detected because of the limited time
baseline of the surveys. This latter effect thus leads to a loss of
sensitivity at large $a$. As a benchmark, Jupiter produces a 12.5 m/s
velocity variation on the Sun with a period of 12 years. The detection
of a Solar System analog requires approximately $k \approx 3$ m/s
precision maintained over a decade window of observing (Paul Butler,
private communication). Since this level of precision (Bernstein et
al. 2003) and this time baseline are the best that are currently
available, the mass and semi-major axis of Jupiter represent a rough
upper limit on detectability. As a general rule, the completeness of
the data set must decline with increasing semi-major axis and
decreasing planet mass. The detection limit can be written in terms of
the reflex velocity $k_{\rm re}$, defined by 
\be 
k_{\rm re} \equiv {m_P \siniobs \over M_\ast^{2/3} 
(1 - \epsilon^2)^{1/2} } \Bigl( {2 \pi G \over P} \Bigr)^{1/3} \, , 
\ee 
which is valid in the limit $m_P \ll M_\ast$. We can scale this
formula to the limit quoted above, i.e., that the detection of Jupiter
itself is near the present day observational threshold. The portion of
the $a-\epsilon$ plane that is accessible to observations is
(conservatively) delimited by the curve $a (1 - \epsilon^2) \le 5
\mu^2$, where $a$ is in AU and $\mu = m_P \siniobs$ is in Jupiter
masses. The radial velocity surveys have 7 year time spans (for the
latter, more complete samples) which implies a limit of about $a \le
3.5$ AU for completeness in semi-major axis. For this value of $a$,
the corresponding mass limit is thus $\mu \ge$ 0.84 $(1 - 
\epsilon^2)^{1/2}$. As a result, for the moderate eccentricities 
observed, $\langle \epsilon \rangle \approx 0.3$, the sample is expected 
to nearly complete out to $a = 3.5 $ AU for $m_P \siniobs \ge 0.8 m_J$, 
and nearly complete at $a$ = 1 AU for $m_P \siniobs \ge 0.43 m_J$. 
At the low end of our planetary IMF, $m_P \sim 0.1 m_J$, the observed 
sample is expected to be missing planets. 

For planets detected with incomplete data sampling, the derived
orbital eccentricities are subject to uncertainties. If a planet has
extremely low eccentricity, then noise in the radial velocity data
stream can mimic the signature of small eccentricities.  On the other
hand, extrasolar planets with the highest eccentricities, say
$\epsilon > 0.8$, may be subject to an additional bias that makes them
hard to detect using available strategies, which are sparsely sampled
in time (due to limited telescope resources). Planets on high
eccentricity orbits spend most of their time out near apstron, where
they produce little radial velocity variation. Unless the system is
observed when the planet is near periastron, it is difficult to
determine the true eccentricity.  Even when such a planet is detected,
the analysis can underestimate the eccentricity until enough data has
been obtained to provide full orbital phase coverage (e.g., see Naef
et al. 2001 and the case of HD 80606).  However, this bias can be
eliminated by sufficient observational coverage and relatively few of
the planets already detected should suffer from this effect (D.
Fischer, private communication).  The eccentricities of multiple
planet systems can also vary with time through secular interactions
(analogous to the secular eccentricity variations of Jupiter and
Saturn). The observed eccentricity distribution of the extrasolar
planets is thus a particular sampling of a larger underlying
distribution. The observed eccentricity values can be either lower or
higher than the mean values sampled over a secular cycle; although
this effect can influence the interpretation of a particular
eccentricity value, it will not affect the overall distribution of
eccentricity of interest here.

\subsection{Additional Evolution of the Orbital Elements} 

The orbital elements of the planets will, in general, continue to
evolve after the end of the simulations presented in the previous
section. In order to compare the theoretical results of this migration
scenario with the orbital elements of observed extrasolar planets,
this additional evolution should be taken into account. In this
section we discuss two physical processes -- continued orbital
evolution driven by the circumstellar disk (with time scale $\sim1$
Myr) and longer term circularization due to interactions of close
planets with the star (with time scale $\sim1$ Gyr).

The simulations end when a planet is ejected or accreted, a collisions
take place, or after a fiducial time span of 1 Myr. In general,
however, the disk will not lose its ability to drive migration at 
exactly the same time that the simulations are stopped. Suppose that 
the disk continues to drive Type II migration (and eccentricity damping) 
over a time span $\Delta t$. The orbital elements will evolve from their 
values $(a_0, \epsilon_0)$ at the end of the numerical simulations to 
new values given by 
\be 
a_f = a_0 \, \, {\rm e}^{-\Delta t / \taudamp} 
\qquad {\rm and} \qquad 
\epsilon_f = \epsilon_0 \, \, {\rm e}^{-\Delta t / \taued} \, . 
\label{eq:diskevolve} 
\ee 
Since the migration time $\taudamp$ and eccentricity damping time 
$\taued$ are determined for a given simulation, the distribution of 
values for the additional migration time $\Delta t$ determines the 
final distribution of orbital elements. 

Unfortunately, the correct choice of the $\Delta t$ distribution is
not known. The numerical experiments begin with the planets already
formed, so the disk has already been around for some time before the
clock starts, and this lead time will vary from system to system. In
fact, within the core accretion scenario of planet formation, theories
often have trouble forming giant planets while the disk retains its
gas (e.g., Kornet et al. 2002 suggest a formation time of about 3
Myr), which leaves little time for migration. The planets with the
largest masses in our numerical simulations lead to the shortest
integration times, but these same planets are expected to have the
longest formation times. Astronomical observations show that
circumstellar disks have lifetimes in the range 3 -- 6 Myr (e.g.,
Haisch et al. 2001), significantly longer than the $\sim1$ Myr time
spans of the integrations. However, this range of observed disk
lifetimes represents the time span over which the disk exhibits
observational signatures. The time over which the disk is sufficiently
massive to drive (Type II) planet migration will be significantly
shorter. In light of these uncertainties, this paper explores a set of
algorithms to account for additional evolution of the orbital elements
by the disk, i.e., a set of simple, but well-defined, distributions
for the remaining migration time $\Delta t$. These distributions and
their effects on the orbital elements are described in the following
subsection.

The system evolution discussed thus far produces orbital elements that
apply to system ages of a few Myr, immediately after planets have
finished forming and the disk has lost its ability to drive migration.
Longer term evolution can also take place. On time scales $\sim1-6$
Gyr characteristic of the stellar ages in observed extrasolar
planetary systems, tidal interactions with the star act to circularize
close orbits. In the absence of other processes, the eccentricity of a
planetary orbit decays with a time scale $\taucirc = {-\epsilon /
{\dot \epsilon}}$. Here we write this time scale in an approximate form 
(see Goldreich and Soter 1966; Hut 1981; Wu and Goldreich 2002): 
\be
\taucirc \approx {2 \over 21} {Q \over k_2} \Bigl( {a^3 \over GM_\ast} 
\Bigr)^{1/2} {m_P \over M_\ast} \Bigl( {a \over R_P} \Bigr)^5 \, 
(1 - \epsilon^2)^{13/2} [ f(\epsilon^2) ]^{-1} \, ,  
\label{eq:circone} 
\ee 
where $Q \approx 10^5 - 10^6$ is the tidal quality factor, $k_2 
\approx 1/2$ is the tidal Love number, and $R_P \approx R_J \approx 7
\times 10^9$ cm is the radius of the planet (notice that this radius
is smaller than that used for the collision cross section in the
simulations because the planets will contract over the longer time
spans considered here). This form includes the essential dependence of
the circularization time scale on eccentricity, where $f(\epsilon^2)$
is a rather complicated function of $\epsilon$ (defined by eqs. 
[9--12] of Hut 1981). Evaluation of this time scale for 
representative values of the parameters indicates that 
\be
\taucirc \approx  \, \, 1 {\rm Gyr} \, \, 
\Bigl[ {a / (0.05 \, {\rm AU}) \, (1 - \epsilon^2) } 
\Bigr]^{13/2} \, [ f(\epsilon^2) ]^{-1} \, . 
\label{eq:circtwo} 
\ee 
As a result, orbits with $a \le 0.05$ AU are expected to be (nearly)
circularized, and orbits with somewhat larger $a$ will experience a
substantial decrease in eccentricity. Although a number of additional
processes can take place (e.g., orbital decay and stellar spinup --
see Lin et al. 2000), the leading order effect is loss of eccentricity
at constant angular momentum (see also Goldreich and Soter 1966; Hut
1981; Wu and Goldreich 2002).  Here we numerically integrate the
(nonlinear) evolution equation for eccentricity over the stellar ages 
$t_\ast$, which are assumed to lie in the range $t_\ast = 1 - 6$ Gyr. 
In the following subsection, we apply this correction to the
theoretical data set in order to compare with observations, although 
only the closest orbits are affected.

\subsection{Comparison of Theory and Observation} 

One way to evaluate the effectiveness of this migration mechanism is
to compare the resulting two-dimensional distribution of orbital
elements in the $a-\epsilon$ plane with those of the observed
extrasolar planets. This subsection presents such a comparison in
systematic fashion.  For all cases discussed here, we use the
log-random IMF, as it closely mirrors the observed mass distribution
of extrasolar planets.  The number of theoretical planets is taken to
be equal to the number in the observed sample, where the theoretical
planets are chosen randomly from the ensemble of simulations.  We then
produce $a-\epsilon$ diagrams for each value of the eccentricity
damping time scale $\taued$ = 0.1, 0.3, 1.0 Myr, and $\taued \to
\infty$. Within this format, the results are presented for a
collection of possible corrections for additional evolution of the
orbital elements and selection effects, as outlined above.

The first comparison in shown in Fig. \ref{fig:aepraw}, which shows
the $a-\epsilon$ plane for observed and theoretical planets, with no
corrections applied for possible additional evolution.  The location
of the observed extrasolar planets in the $a-\epsilon$ plane are
marked by the star symbols, whereas the results of the theoretical
simulations are marked by open squares. The four panels show the
results of the four eccentricity damping time scales.  The theoretical
distribution of planets moves to lower values of eccentricity and to
lower values of semi-major axis as the eccentricity damping time scale
decreases. The lower $\epsilon$ values are a direct result of the
increased effectiveness of eccentricity damping. The lower values of
$a$ occur because the increased eccentricity damping keeps the planets
stable for longer times and the disk has more time to move planets
inward. This figure suggests that the simulations with no eccentricity
damping ($\taued \to \infty$; lower right panel) produce too many high
eccentricity planets compared to the observational sample, whereas the
simulations with $\taued$ = 0.1 Myr (upper left panel) tend to produce
planetary orbits with too little eccentricity. Nonetheless, the
zeroeth order result of this comparison is that both the observed
planets and the theoretical simulations fill most of the $a-\epsilon$
plane shown here (except for the case $\taued \to \infty$, which
ejects planets before they move far enough inward). Nonetheless, some
differences appear, and we need to explore whether or not the
corrections for additional evolution described above act to bring the
observational and theoretical samples into better agreement.

The discussion of Section 4.1 suggests that the observed sample is
likely to be incomplete for planet masses below 0.5 $m_J$, with the
level of incompleteness increasing with semi-major axis $a$. In other
words, the sample is incomplete for small values of the reflex
velocity $k_{\rm re}$. To determine the importance of this issue on
our assessment of this migration mechanism, we present the same set of
$a-\epsilon$ diagrams with a $k_{\rm re}$ cut applied; specifically,
all planets in the theoretical sample with $k_{\rm re} \le 3$ m/s have
been removed from consideration. The result is shown in Fig.
\ref{fig:aepmass}. No other corrections for additional evolution have
been applied to the theoretical sample.  Comparison of Figs.
\ref{fig:aepraw} and \ref{fig:aepmass} indicates that mass/$k_{\rm
re}$ incompleteness has only a modest effect on comparisons of the
$a-\epsilon$ plane. This result makes sense because the mass range in
question, roughly $m_P \le 0.5 m_J$, represents about one fourth of
the starting mass range, but relatively more of the low mass planets
are ejected or accreted (see Fig.  \ref{fig:loghist} and Table 2). As
a result, only 10 -- 15 percent of the surviving planets fall in this
low mass range.

In order to take into account additional evolution of the orbital
elements beyond the end of the simulations due to the surviving
circumstellar disk, we apply corrections according to equation
(\ref{eq:diskevolve}). As discussed above, the distribution of
additional migration time $\Delta t$ is not well determined. As a
result, we explore different algorithms for continued evolution. In
the first case, we assume that the disk is able to drive migration
beyond the end of the numerical simulations for an additional time
given by $\Delta t$ = $\delta t - t_{sim}$, where $t_{sim}$ is the
time at the end of the simulation, $\delta t$ is a random time scale
in the range 0 -- 1 Myr (where negative values of $\Delta t$ are set
to zero, i.e., no additional evolution). This is the same algorithm
used in the previous study of this migration scenario in AL2003.  The
resulting $a-\epsilon$ diagrams are shown in Fig. \ref{fig:aepold} for
different choices of eccentricity damping time scale. The continued
migration moves the points to lower values of both $a$ and $\epsilon$,
and the random time element tends to spread the distributions, 
although the effect is relatively small (except for the case with 
no eccentricity damping). Nonetheless, this 
correction acts to bring the theoretical and observational
distributions into closer agreement (although the distribution of 
$\Delta t$ applied here is not unique). 

An alternate assumption for additional evolution is that the disk has
a remaining lifetime $\Delta t$ that is random and independent of the
previous evolution. Keep in mind that $\Delta t$ is the time over
which the disk has enough mass to change the orbital elements of any
remaining planets; the disk may exhibit observational signatures over
longer times. Fig. \ref{fig:aepran3} shows the resulting $a-\epsilon$
diagrams for random disk lifetimes in the range $\Delta t$ = 0 -- 0.3
Myr. Fig. \ref{fig:aepran5} shows the $a-\epsilon$ diagrams for random
disk lifetimes in the somewhat longer range $\Delta t$ = 0 -- 0.5 Myr.
These results are much the same as for the previous algorithm
illustrated in Fig. \ref{fig:aepold}. Any similar model of continued
disk evolution will move the theoretical points to smaller values of
$(a,\epsilon)$ and will spread the distributions. The magnitude of the
effect (the mean value of the $\Delta t$ distribution) matters more
than the particular choice of algorithm (which sets the shape of the
distribution). Inspection of Figs. \ref{fig:aepold} --
\ref{fig:aepran5} suggests that a mean value $\langle \Delta t \rangle
\approx 0.3 - 0.5$ Myr is needed to provide reasonable agreement with
observations. Notice that the model allows for some interplay between
this time scale and the eccentricity damping time scale -- for larger
$\taued$, less additional evolutionary time $\langle \Delta t \rangle$ 
is indicated.

Next we consider corrections to the orbital elements due to the longer
term process of tidal circularization by the central star. To account
for this effect, we integrate the differential equation ${\dot
\epsilon}/\epsilon = - \taucirc^{-1}$, with the circularization time
scale given by equations (\ref{eq:circone}, \ref{eq:circtwo}). The
assumed system lifetimes are assumed to be randomly distributed and
lie in the range 1 -- 6 Gyr, similar to the stellar ages in the
observed sample. The resulting $a-\epsilon$ diagrams are shown in
Fig. \ref{fig:circular}. The inclusion of this circularization
processes cleans up an important discrepancy between the theoretical
and observed orbital elements, namely the observed lack of short
period planets (small $a$) with substantial eccentricities.

Finally, in Fig. \ref{fig:allcuts}, we present a set of $a-\epsilon$
diagrams with all of the corrections applied: a reflex velocity cut
such that only planets with $k_{\rm re} > 3$ m/s remain, continued
migration with remaining disk lifetimes given by the algorithm
depicted in Fig. \ref{fig:aepold}, and the circularization
correction. The resulting theoretical distributions of $a$ and
$\epsilon$ are in reasonable agreement with those of the observed
sample of extrasolar planets.  The eccentricity damping time scale
$\taued$ = 0.3 Myr (i.e., $\taued = \taudamp$) provides the best fit,
although all three cases with $\taued$ in the range 0.3 -- 3 Myr are
in the right ballpark.

\section{Conclusion} 

\begin{figure} 
{\epsscale{1.0} \plotone{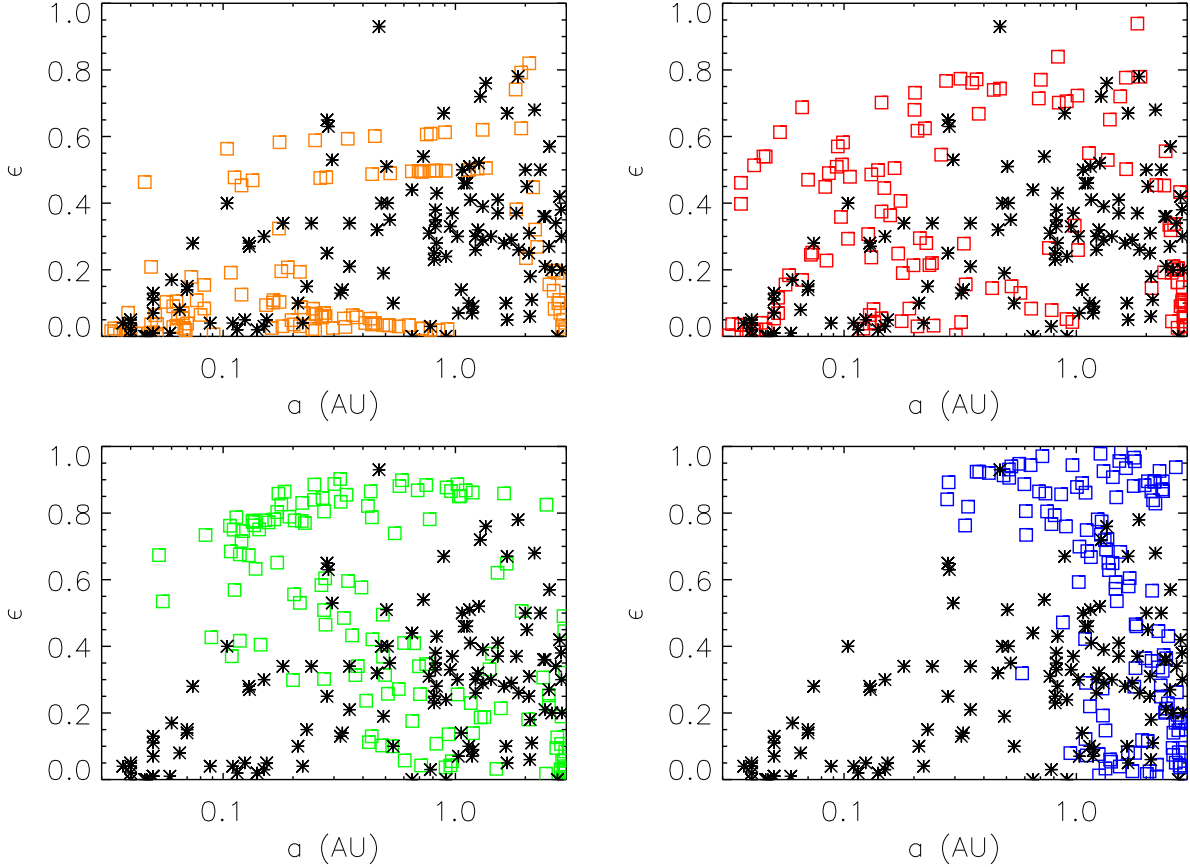}} 
\caption{The $a-\epsilon$ plane for observed and theoretical planets,
where no corrections for additional evolution have been applied to the
theoretical sample. This diagram shows the semi-major axes $a$ and
eccentricities $\epsilon$ for the observed extrasolar planets as
stars. The results of the theoretical simulations are shown as open
squares. All of the theoretical simulations use the log-random
IMF. The four panels correspond to different choices of the
eccentricity damping time scale: $\taued$ = 0.1 Myr (upper left),
$\taued$ = 0.3 Myr (upper right), $\taued$ = 1.0 Myr (lower left), and
$\taued \to \infty$ (lower right). }  
\label{fig:aepraw}
\end{figure}

\begin{figure} 
{\epsscale{1.0} \plotone{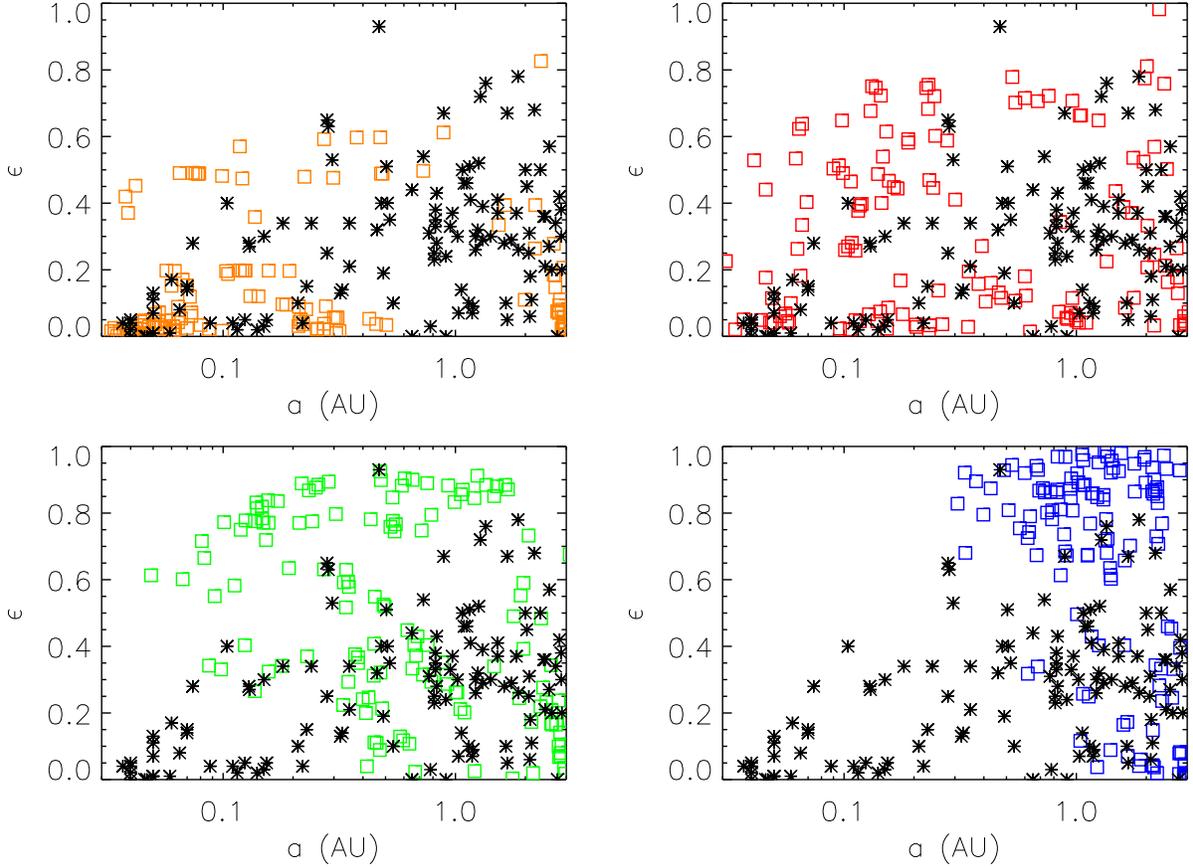}}  
\caption{The $a-\epsilon$ plane for observed and theoretical planets,
where the theoretical sample starts with a log-random IMF and has been
subjected to a cut in reflex velocity $k_{\rm re}$ at 3 m/s. No corrections 
for additional evolution have been applied to the theoretical sample. 
This diagram shows the semi-major axes $a$ and eccentricities $\epsilon$ 
for the observed extrasolar planets as stars. The results of the
theoretical simulations are shown as open squares. The four panels
correspond to different choices of the eccentricity damping time
scale: $\taued$ = 0.1 Myr (upper left), $\taued$ = 0.3 Myr (upper
right), $\taued$ = 1.0 Myr (lower left), and $\taued \to \infty$
(lower right). }  
\label{fig:aepmass} 
\end{figure}

\begin{figure} 
{\epsscale{1.0} \plotone{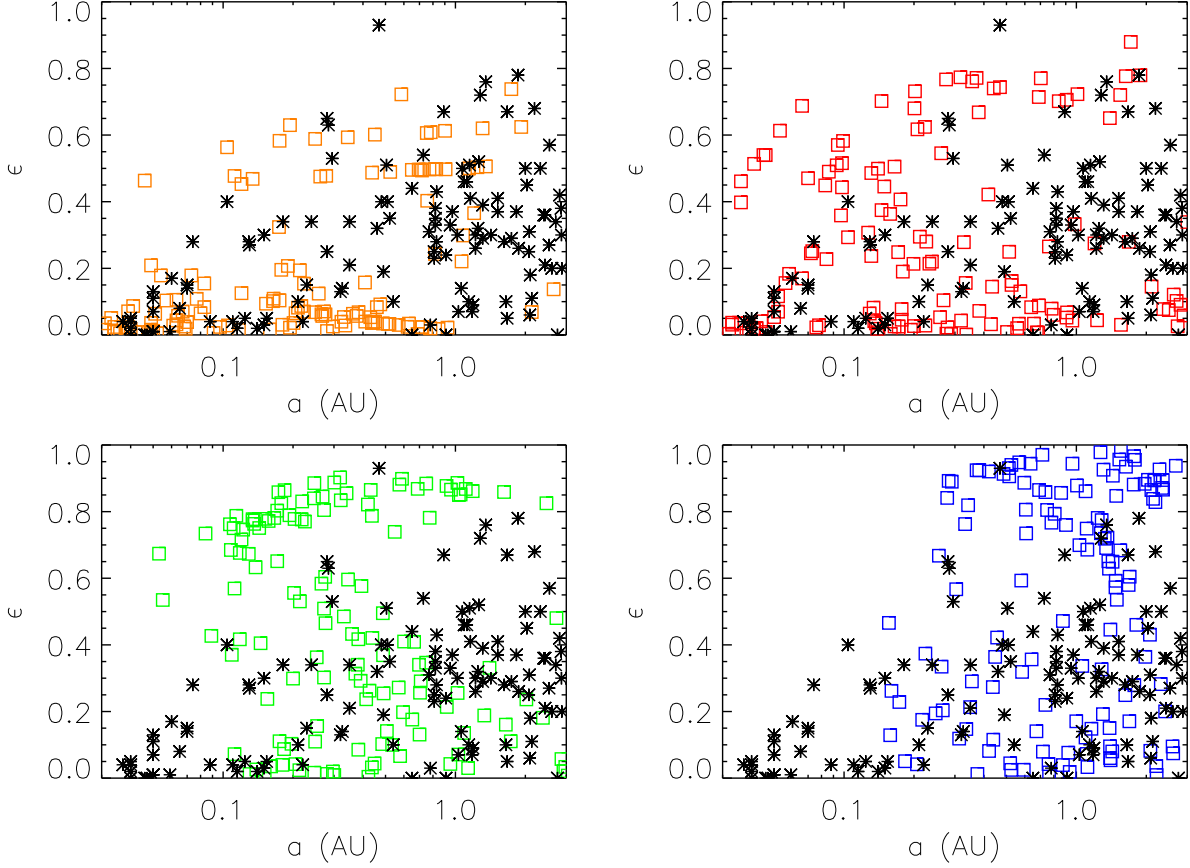}}   
\caption{The $a-\epsilon$ plane for observed and theoretical planets,
where the theoretical sample starts with a log-random IMF and has been
corrected for additional orbital evolution (first alogrithm).  Here
the disk is assumed to exist beyond the end of the numerical
simulations for an additional time given by $\Delta t$ = $\delta t -
t_{sim}$, where $t_{sim}$ is the time at the end of the simulation,
$\delta t$ is a random time scale in the range 0 -- 1 Myr, and
negative values are set to zero.  This diagram shows the semi-major
axes $a$ and eccentricities $\epsilon$ for the observed extrasolar
planets as stars. The results of the theoretical simulations are shown
as open squares. The four panels correspond to different choices of
the eccentricity damping time scale: $\taued$ = 0.1 Myr (upper left),
$\taued$ = 0.3 Myr (upper right), $\taued$ = 1.0 Myr (lower left), and
$\taued \to \infty$ (lower right). }  
\label{fig:aepold} 
\end{figure}

\begin{figure} 
{\epsscale{1.0} \plotone{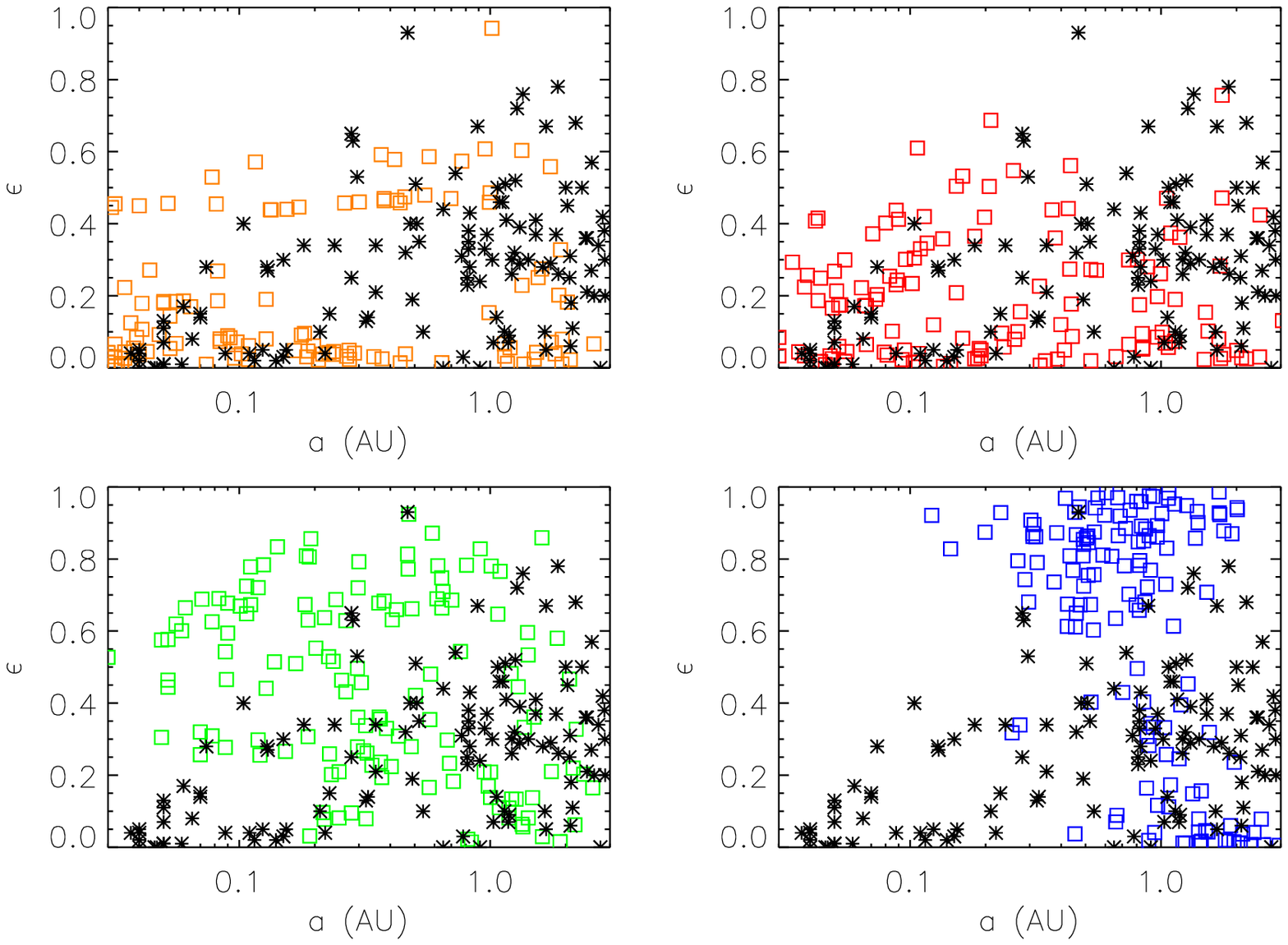}}   
\caption{The $a-\epsilon$ plane for observed and theoretical planets,
where the theoretical sample starts with a log-random IMF and has been
corrected for additional orbital evolution (second algorithm). The disk 
is assumed to exist beyond the end of the numerical simulations for an
additional time $\Delta t$, which is chosen randomly from the range 0
-- 0.3 Myr.  This diagram shows the semi-major axes $a$ and
eccentricities $\epsilon$ for the observed extrasolar planets as
stars. The results of the theoretical simulations are shown as open
squares. The four panels correspond to different choices of the
eccentricity damping time scale: $\taued$ = 0.1 Myr (upper left),
$\taued$ = 0.3 Myr (upper right), $\taued$ = 1.0 Myr (lower left), and
$\taued \to \infty$ (lower right). }  
\label{fig:aepran3} 
\end{figure}

\begin{figure} 
{\epsscale{1.0} \plotone{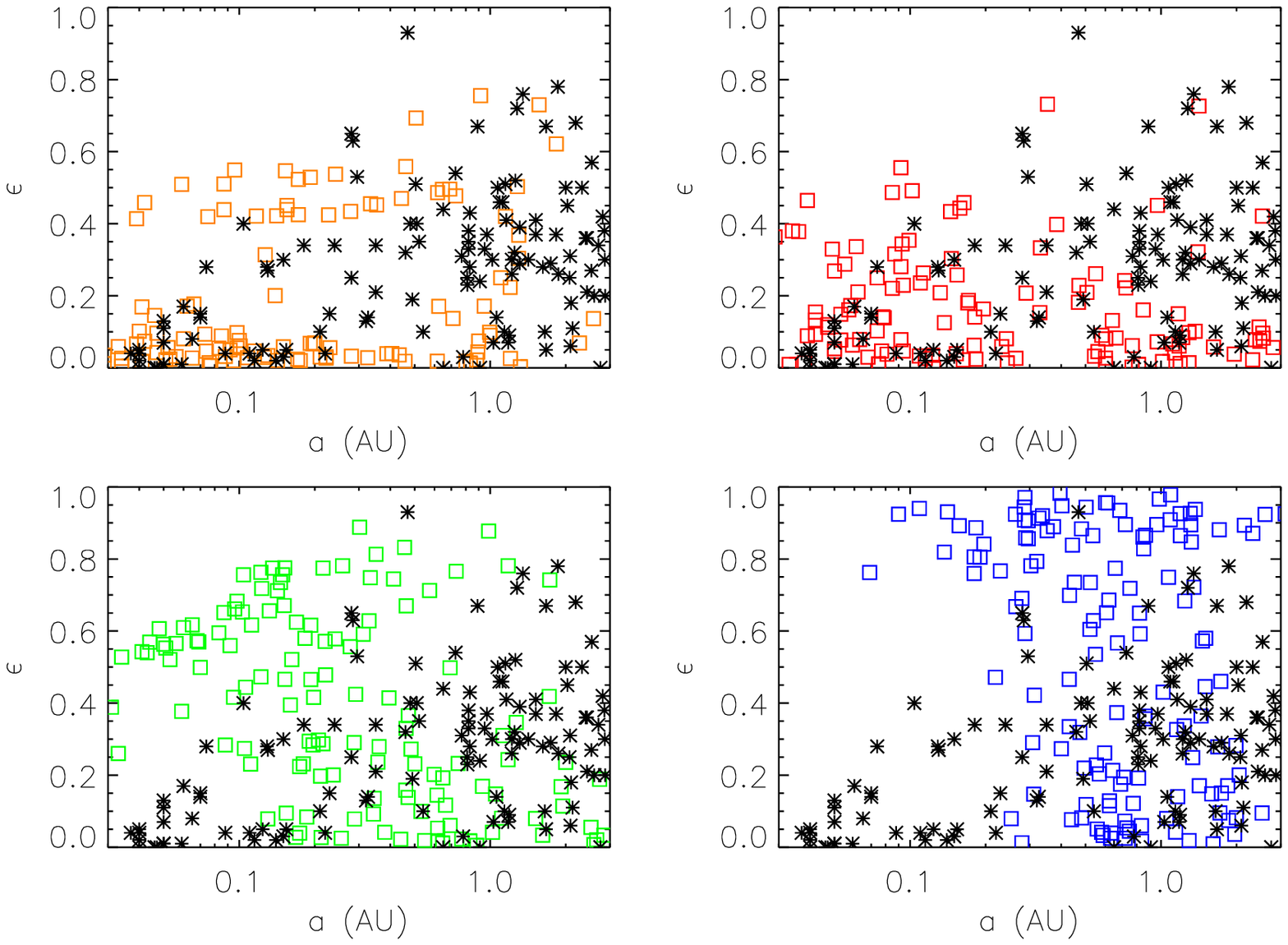}}   
\caption{The $a-\epsilon$ plane for observed and theoretical planets,
where the theoretical sample starts with a log-random IMF and has been
corrected for additional orbital evolution (third alogorithm). The
disk is assumed to exist beyond the end of the numerical simulations
for an additional time $\Delta t$, which is chosen randomly from the
range 0 -- 0.5 Myr.  This diagram shows the semi-major axes $a$ and
eccentricities $\epsilon$ for the observed extrasolar planets as
stars. The results of the theoretical simulations are shown as open
squares. The four panels correspond to different choices of the
eccentricity damping time scale: $\taued$ = 0.1 Myr (upper left),
$\taued$ = 0.3 Myr (upper right), $\taued$ = 1.0 Myr (lower left), 
and $\taued \to \infty$ (lower right). }  
\label{fig:aepran5} 
\end{figure}

\begin{figure} 
{\epsscale{1.0} \plotone{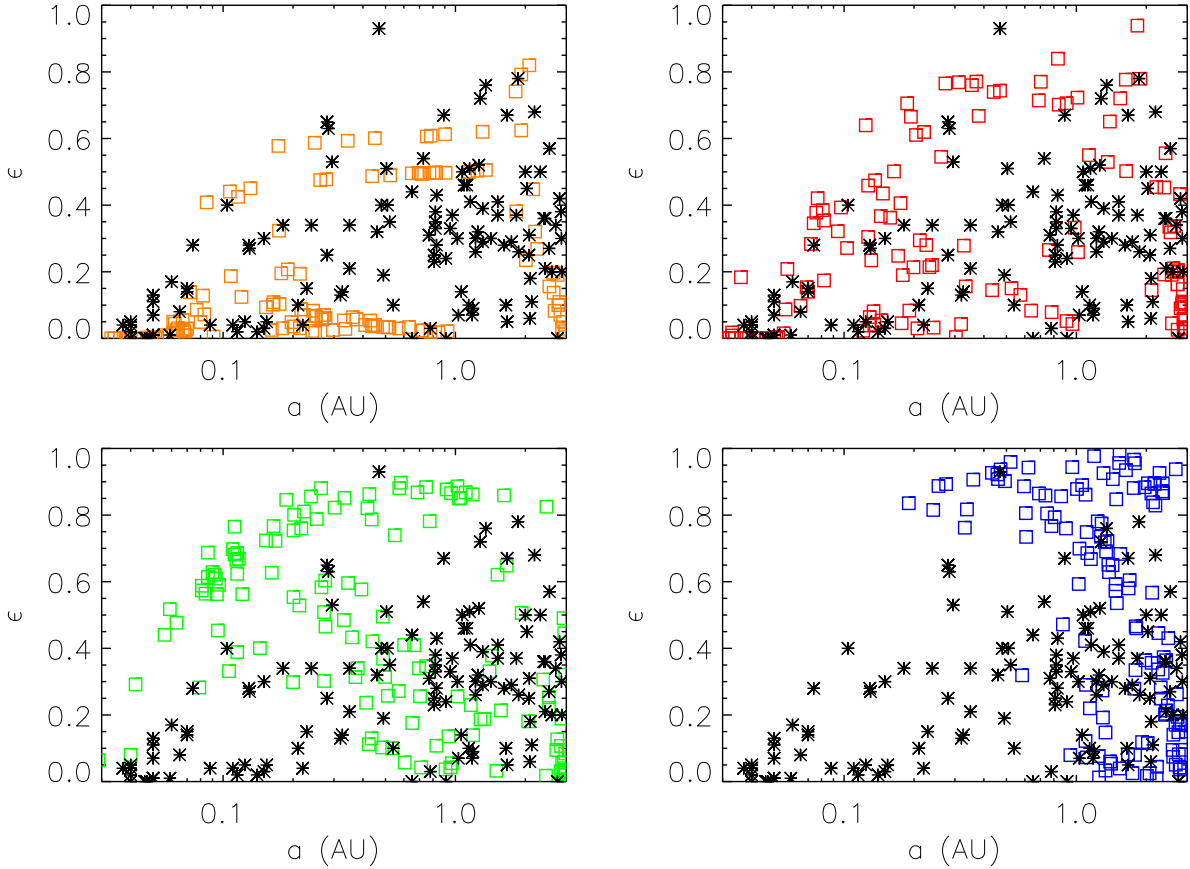}}   
\caption{The $a-\epsilon$ plane for observed and theoretical planets,
where the theoretical sample starts with a log-random IMF and has been
corrected for tidal circularization over the stellar lifetime, which
is assumed to lie in the range 1 -- 6 Gyr.  This diagram shows the
semi-major axes $a$ and eccentricities $\epsilon$ for the observed
extrasolar planets as stars; results of the theoretical simulations 
are shown as open squares. The four panels correspond to different
choices of the eccentricity damping time scale: $\taued$ = 0.1 Myr
(upper left), $\taued$ = 0.3 Myr (upper right), $\taued$ = 1.0 Myr
(lower left), and $\taued \to \infty$ (lower right). }
\label{fig:circular} 
\end{figure}

\begin{figure} 
{\epsscale{1.0} \plotone{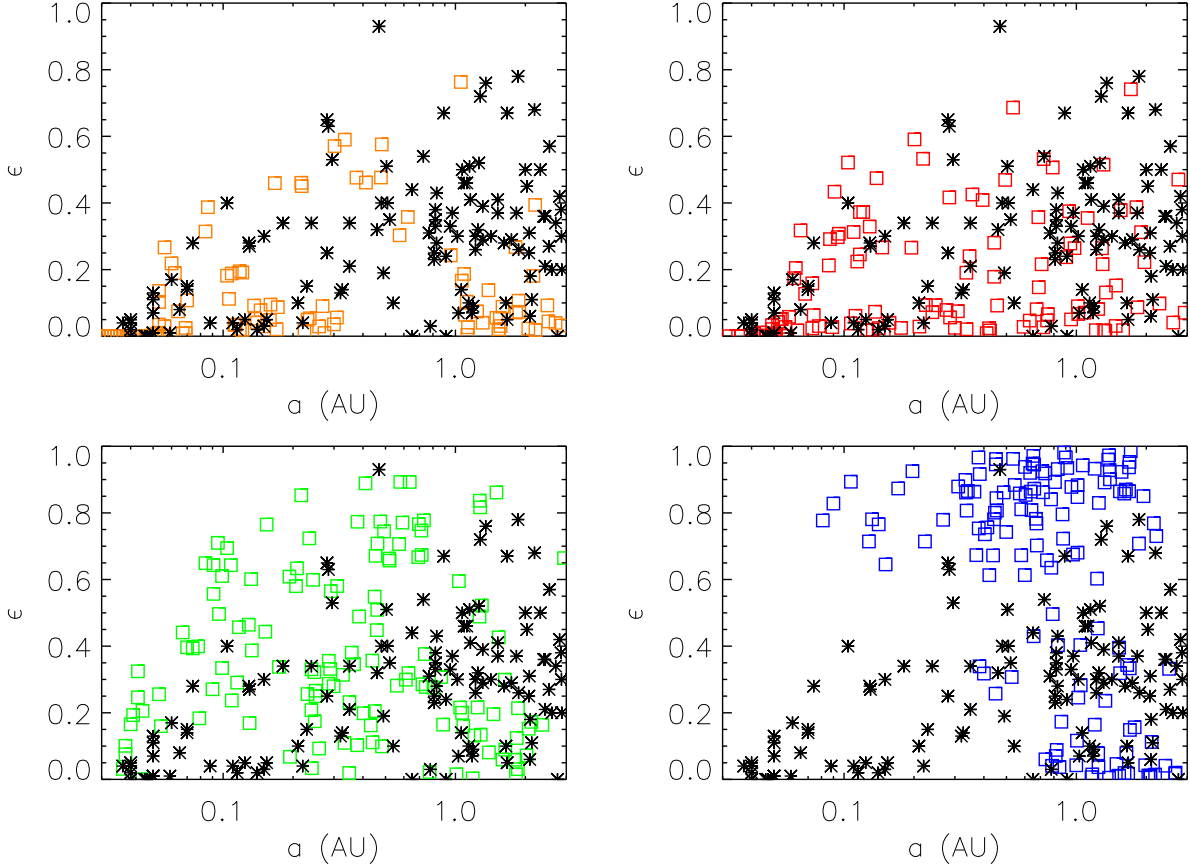}}   
\caption{The $a-\epsilon$ plane for observed and theoretical planets
using corrections for both continued disk evolution and tidal
circularization. The theoretical sample starts with a log-random IMF,
but a reflex velocity cut $k_{\rm re} > 3$ m/s has been applied to the 
surviving planets. The disk is assumed to continue driving planet migration 
beyond the end of the simulations for a random time interval in the
range 0 -- 0.3 Myr. Tidal circularization is assumed to continue for
a stellar lifetime, taken to be a random time interval in range 1 -- 6
Gyr. This diagram shows the semi-major axes $a$ and eccentricities
$\epsilon$ for the observed extrasolar planets as stars; results of
the theoretical simulations are shown as open squares. The four panels
correspond to different choices of the eccentricity damping time
scale: $\taued$ = 0.1 Myr (upper left), $\taued$ = 0.3 Myr (upper
right), $\taued$ = 1.0 Myr (lower left), and $\taued \to \infty$
(lower right). }  
\label{fig:allcuts} 
\end{figure}

This paper explores a migration scenario in which multiple giant
planets are driven inward through the action of tidal torques in a
circumstellar disk. In this case, the outer planet interacts with the
disk, which drains energy and angular momentum away from the planetary
orbit. As the outer planet migrates inward, it eventually becomes
close enough to the interior planet to force it inward and to drive
eccentricity growth with increasingly violent interactions. Such
systems are generally not stable in the long term and adjust
themselves to stability by ejecting a planet, accreting a planet onto
the central star, or by having the two planets collide. The surviving
planet is left on an eccentric orbit of varying semi-major axis,
roughly consistent with the orbits of observed extrasolar planets. On
longer time scales, tidal interactions with the central star act to
circularize the orbits of the closet planets.  We have presented a
comprehensive, but not exhaustive, exploration of parameter space for
this migration scenario. Our main results can be summarized as
follows:

[1] This migration scenario results in a wide variety of final systems
with a broad distribution of orbital elements. In particular, this
migration scenario can fill essentially the entire $a-\epsilon$ plane
for semi-major axes $a$ smaller than the initial values. The observed
extrasolar planets have orbital elements that fill the $a-\epsilon$
plane in roughly the same way (see Figs. \ref{fig:aepraw} --
\ref{fig:allcuts}). When the theoretical ensemble of planets is
corrected for additional orbital evolution due to interactions with
the circumstellar disk (on $\sim1$ Myr time scales) and tidal
interactions with the central star (on $\sim1$ Gyr time scales), the
resulting distributions of theoretical orbital elements are in
reasonable agreement with those of the observed sample of extrasolar
planets.

\begin{figure} 
{\centerline {\epsscale{1.0} \plotone{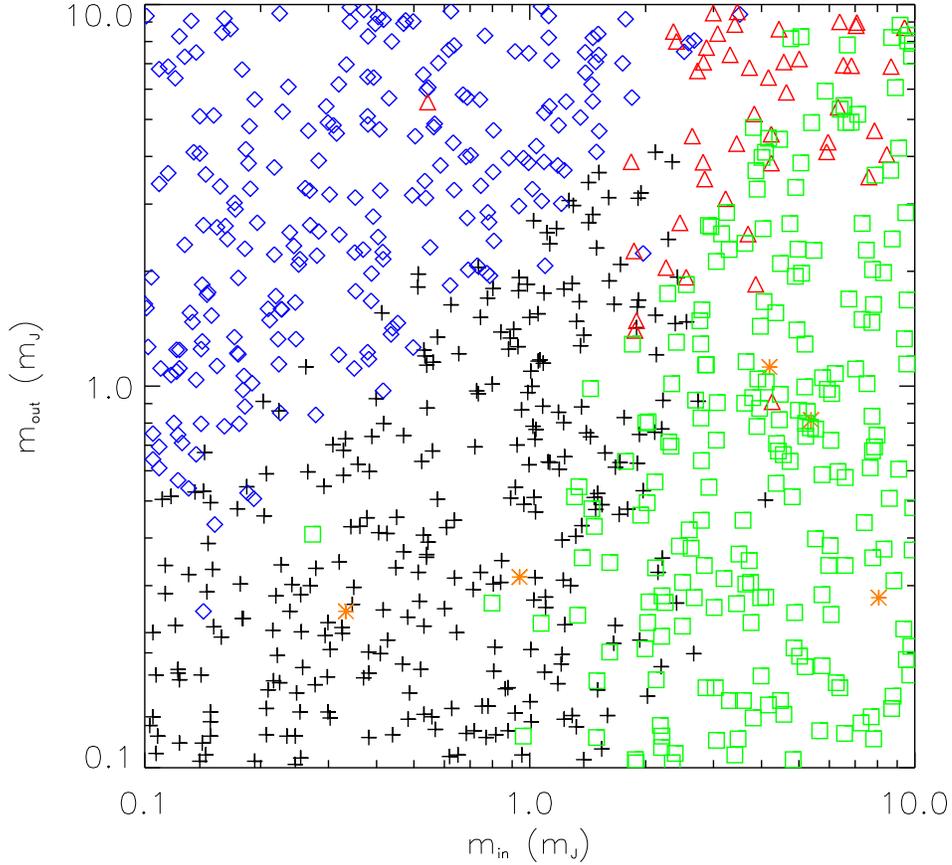} } } 
\vskip -5.0truein
\caption{End states as a function of the planetary masses for
eccentricity damping time scale $\taued$ = 1 Myr.  Each symbol in this
figure represents the outcome of a simulation with the mass of the
outer planet plotted as a function of the mass of the inner planet.
All of the simulations depicted here use the log-random IMF, the
standard starting configuration in which the outer planet begins
outside the 3:1 resonance, and eccentricity damping time $\taued$ = 1
Myr.  The different symbols correspond to different outcomes: open
blue diamonds represent accretion of the inner planet, open green
squares denote ejection of the outer planet, open red triangles
represent ejection of the inner planet, crosses denote survival of
both planets, and orange star symbols represent accretion of the outer
planet.}  
\label{fig:massplot} 
\end{figure}

\begin{figure} 
{\centerline {\epsscale{1.0} \plotone{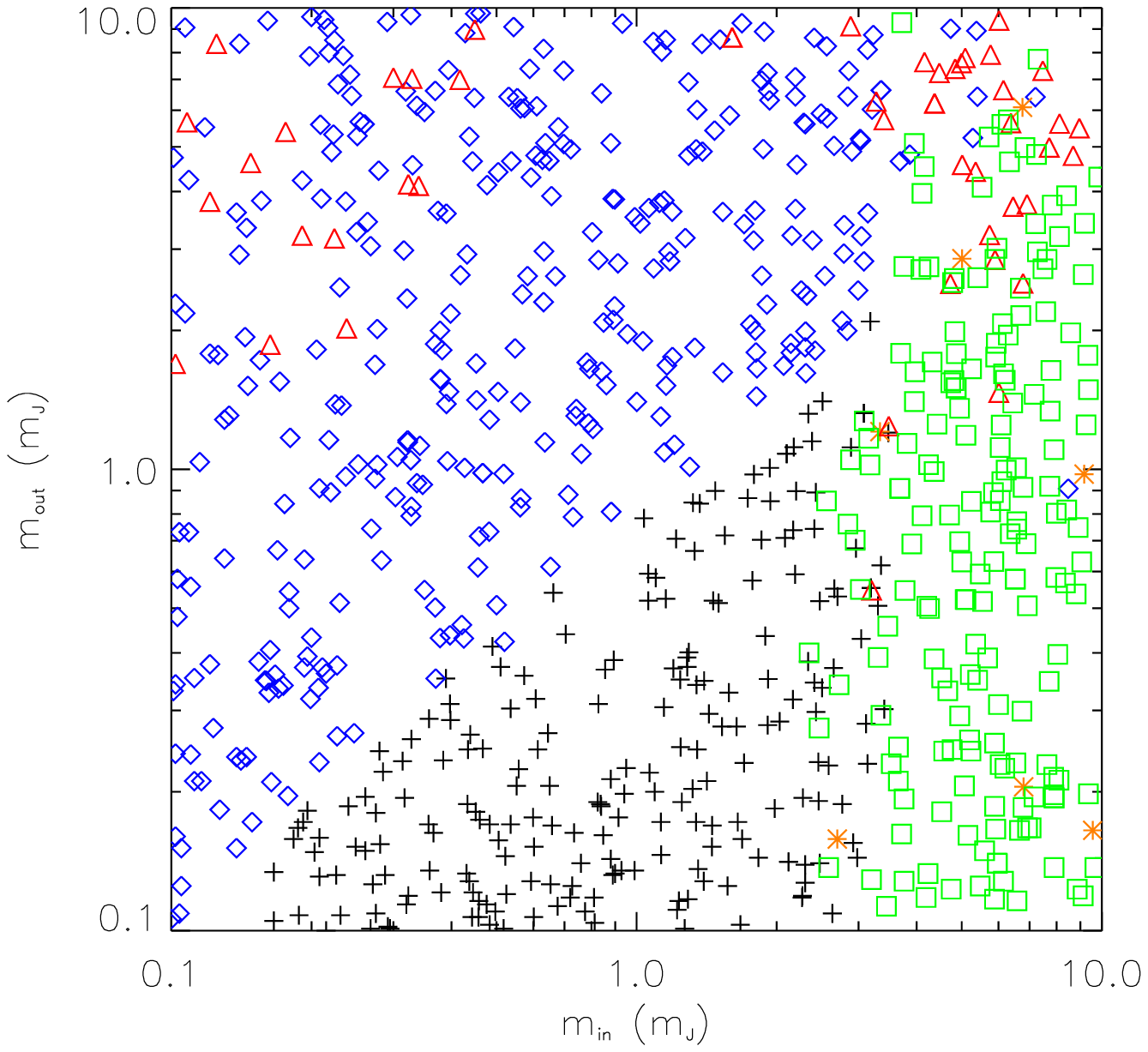} } } 
\vskip -5.0truein
\caption{End states as a function of the planetary masses for 
eccentricity damping time scale $\taued$ = 0.1 Myr (compare with 
Fig. \ref{fig:massplot}). Each symbol in this figure represents the
outcome of a simulation with the mass of the outer planet plotted as a
function of the mass of the inner planet.  All of the simulations
depicted here use the log-random IMF, the standard starting
configuration in which the outer planet begins outside the 3:1
resonance, and eccentricity damping time $\taued$ = 0.1 Myr.  The
different symbols correspond to different outcomes: open blue diamonds
represent accretion of the inner planet, open green squares denote
ejection of the outer planet, open red triangles represent ejection of
the inner planet, crosses denote survival of both planets, and orange
star symbols represent accretion of the outer planet.}
\label{fig:massplotalt} 
\end{figure}

\begin{figure} 
{\hskip -0.5truein {\epsscale{1.10} \plotone{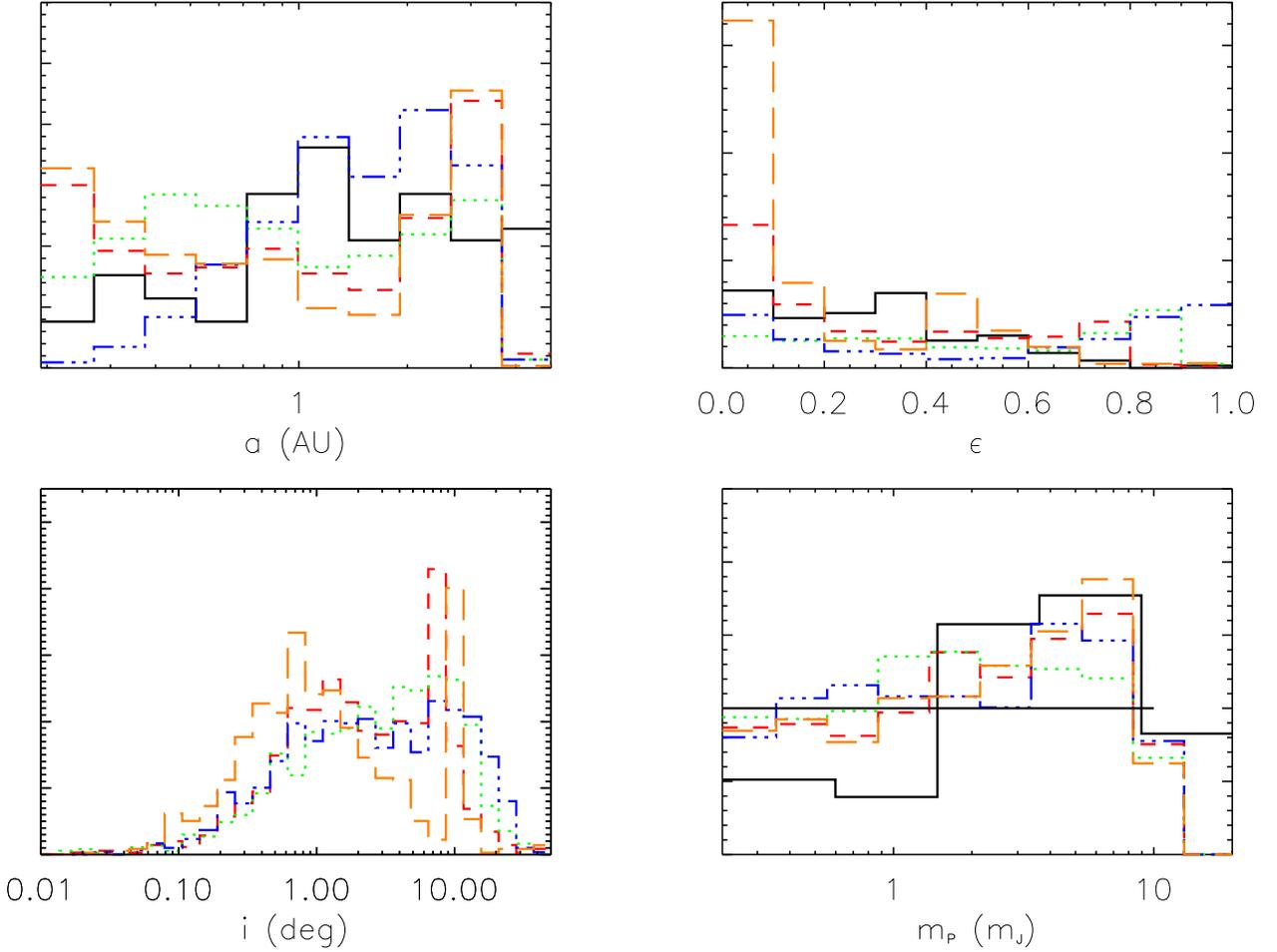} } }   
\caption{Normalized histograms of the orbital elements of surviving
planets for a logarithmic (random) planetary IMF. The upper left panel
shows the distributions of semi-major axis for the observed planets
(solid curve) and theoretical simulations with varying time scales for
eccentricity damping: long dashes ($\taued$ = 0.1 Myr), regular dashes 
($\taued$ = 0.3 Myr), dots ($\taued$ = 1 Myr), and double dot-dashes
($\taued \to \infty$). Similarly, the upper right panel shows the
distributions of eccentricities and the lower left panel shows the
distributions of orbital inclination angles. The lower right panel
shows the distributions of masses, where the solid histogram shows 
the distribution of observed planets and the solid horizontal line 
shows starting log-random distribution; note that the log-random IMF 
for the simulations cuts off at 10 $m_J$. } 
\label{fig:loghist} 
\end{figure}

[2] Planets of smaller mass tend to be the ones that are ejected or
accreted (see Figs. \ref{fig:massplot} and \ref{fig:massplotalt}).
The mass distribution of the observed planetary sample is roughly
log-random, with a moderate deficit of planets at the low mass end
(Fig. \ref{fig:loghist}). This shape is consistent with planets being
formed with (roughly) a log-random mass distribution and the lower end
of the mass function being depleted through planet-planet scattering,
as produced by this migration mechanism. Keep in mind, however, that
the lower end of the mass distribution also suffers from selection
effects (Tabachnik and Tremaine 2002), which must be sorted out before
definitive conclusions can be made (see Section 4.1).

[3] The mutual gravitational interactions of the planets are highly
effective at increasing orbital eccentricities. The general trend is
for planet-planet scattering to produce orbital eccentricities that
are somewhat larger than those observed in the current sample of
extrasolar planets.  As a result, real solar systems must either
provide sufficient eccentricity damping as suggested by numerical
simulations of planet-disk interactions (e.g., Kley et al. 2004;
Nelson et al. 2000), contain only single planets, or provide a
mechanism to keep multiple planets sufficiently separated. The
eccentricity damping time scale that provides the best fit to the
observations lies in the range $\taued$ = 0.1 -- 1 Myr for a viscous
damping time scale of $\taudamp$ = 0.3 Myr, i.e., the ratio
$\taued/\taudamp$ = 1/3 -- 3.

%Notice that if planet-disk interactions were to excite orbital
%eccentricity as suggested in recent work (e.g., Ogilvie and Lubow
%2003; Sari and Goldreich 2004), then the presence of multiple planets
%would lead to much higher eccentricities than those of the
%observational sample.

\begin{figure} 
{\hskip -0.5truein {\epsscale{1.10} \plotone{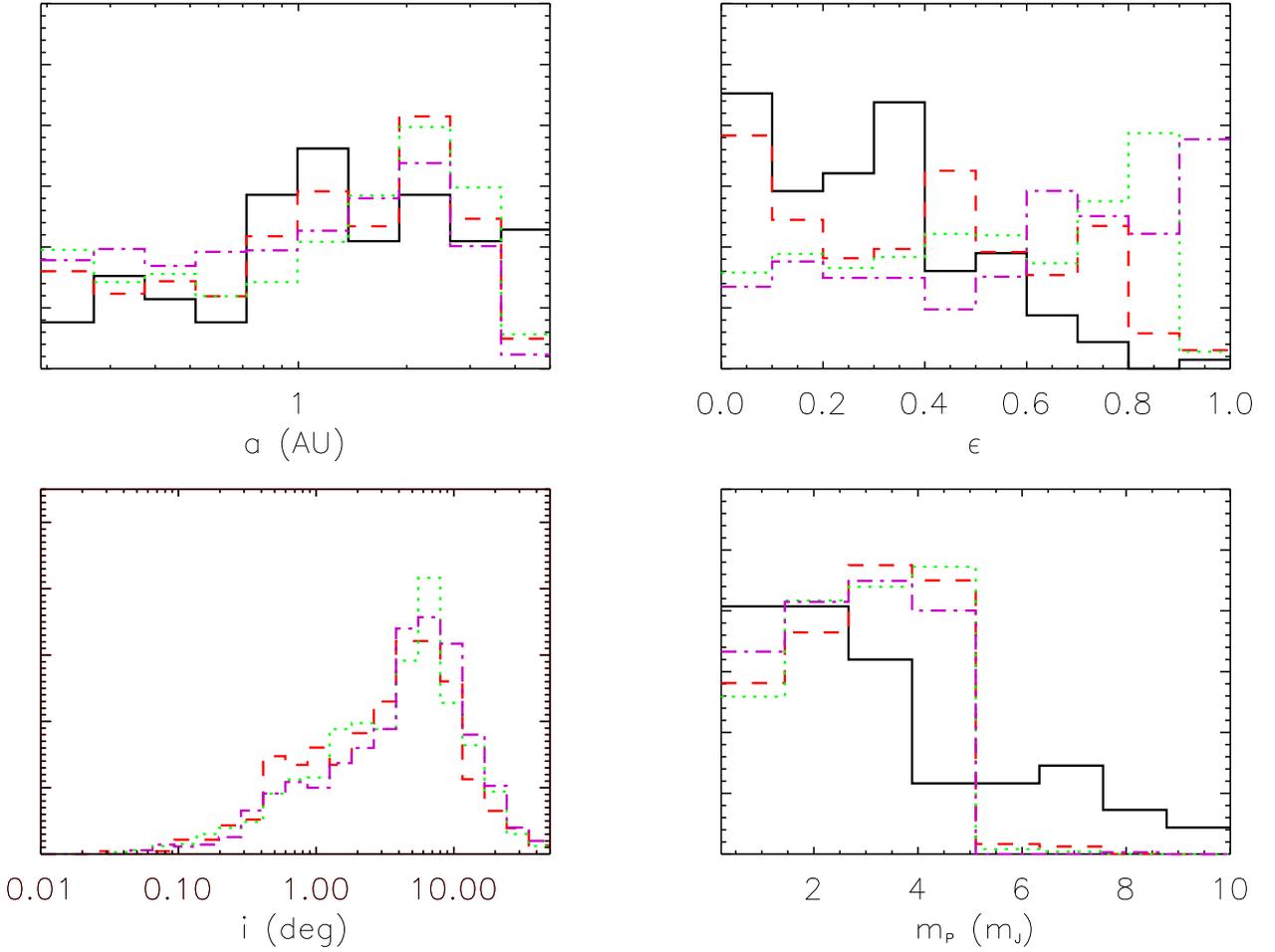} } } 
\caption{Normalized histograms of the orbital elements of surviving
planets for a linear (random) planetary IMF. The upper left panel
shows the distributions of semi-major axis for the observed planets
(solid curve) and theoretical simulations with varying time scales for
eccentricity damping: dashes ($\taued$ = 0.3 Myr), dots ($\taued$ = 1
Myr), and dot-dashes ($\taued$ = 3 Myr).  Similarly, the upper right
panel shows the distributions of eccentricities and the lower left panel
shows the distributions of orbital inclination angles. The lower right 
panel shows the distributions of masses, where the mass distribution 
of the observed planets (solid curve) is included for comparison; 
note that the random IMF for the simulations cuts off at 5 $m_J$. }  
\label{fig:linhist}
\end{figure}

[4] The inclination angles of the planetary orbits are excited with a
well-defined distribution centered of $\Delta i \approx$ 5 degrees
(Figs. \ref{fig:linhist} and \ref{fig:loghist}). For the end states
of this migration mechanism, the inclination angle excitation is
correlated with the excitation of orbital eccentricity (Table 6). 

\begin{figure} 
{\hskip -0.5truein \plotone{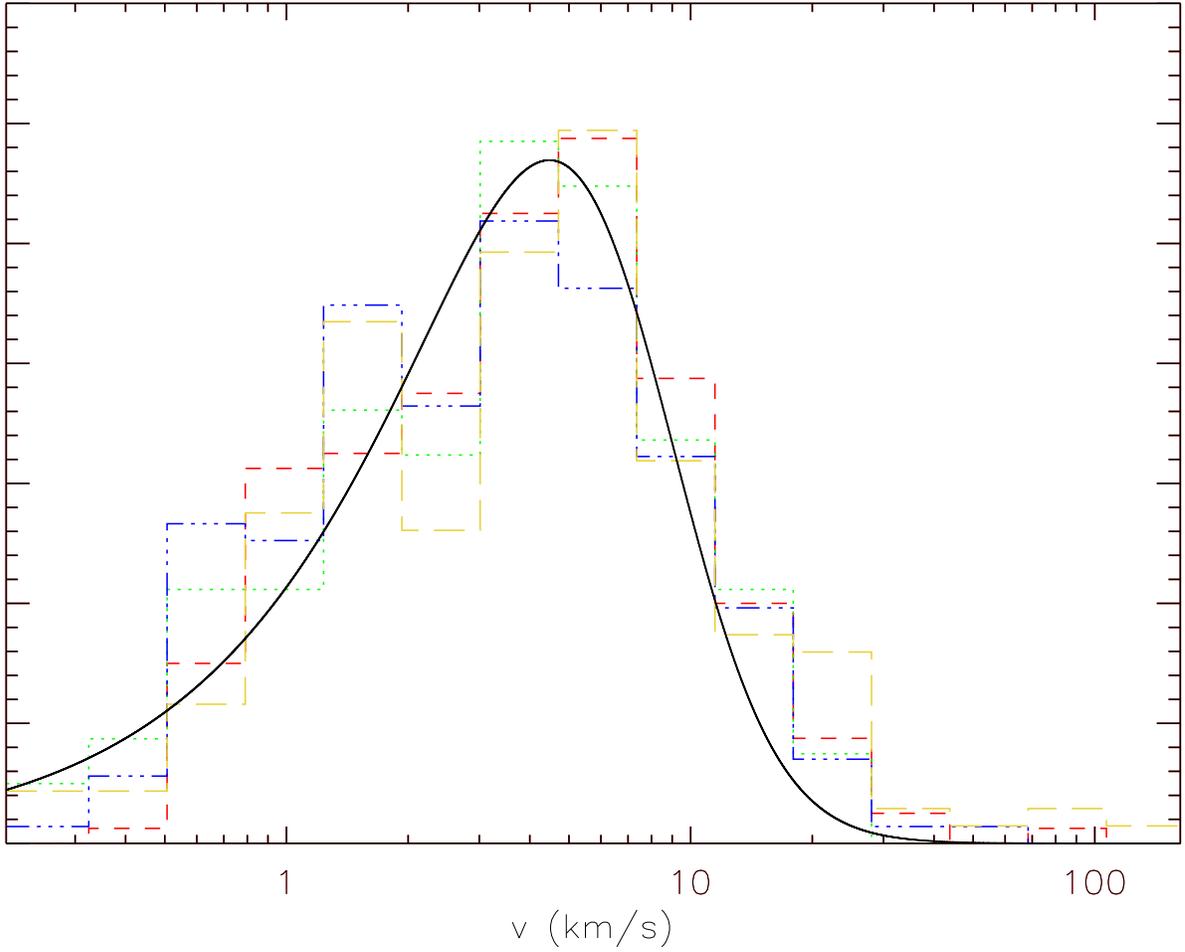}} 
\caption{Distribution of ejection velocities for planets that are
ejected during the epoch of migration. The distribution is shown for
the log-random planetary IMF and for four values of the eccentricity
damping time scale: $\taued$ = 0.1 Myr (yellow long-dashed curve),
$\taued$ = 0.3 Myr (red short-dashed curve), $\taued$ = 1.0 Myr (green
dotted curve), and $\taued \to \infty$ (blue dot-dashed curve). The
four distributions are normalized to the same value, with the vertical
scale arbitrary. The smooth solid curve shows the (normalized) analytic
approximation to the distribution of ejection speeds (as derived in the
text).  All speeds are given in km/s. } 
\label{fig:veject}
\end{figure}

[5] This migration scenario leads to a large number of ejected
planets; specifically, about one-third to one-half of the simulated
systems eject a planet. The distribution of ejection speeds is broad,
with a peak near 5 km/s and a long tail toward higher speeds (Fig.
\ref{fig:veject}). The functional form of the distribution of ejection
speeds can be understood in terms of the simple physical argument
given in Section 3.5. More than 90 percent of the exiled planets are
predicted to attain ejection speeds greater than 1 km/s, the speed
required for planets to (immediately) escape their birth aggregate.
As a result, typical stellar birth clusters (with $N_\ast \approx
300$) are expected to contain only $\sim12$ free floating planets at 
a given time due to this migration mechanism. 

This migration scenario produces a full distribution of orbital
elements for the surviving planets and is in reasonable agreement with
observations. In order for this mechanism to be successful, the
systems must have a number of properties, and it is useful to
summarize them here: The planets that end up in the currently observed
region of the $a-\epsilon$ plane are assumed to have formed in disk
annulus $r = 3 - 7$ AU, roughly where Jupiter lives in our solar
system. The disk must be able to sustain Type II migration torques
over time scales $\sim1$ Myr; the disk must maintain more mass (in
gas) than its planets over this time scale, which is comparable to the
time required for giant planets to form through the core accretion
mechanism. Some system-to-system variation in this migration time
scale is also indicated. As mentioned above, disk signatures are
observed over longer times (several Myr), although the disks do not
necessarily maintain enough mass to drive migration over this longer
time. The torques must be large enough so that $|a/{\dot a}| \approx$
0.3 Myr, which is equivalent to having a viscosity parameter $\alpha
\sim 10^{-3}$. The disk must damp orbital eccentricity of the outer
planet with a damping time scale in the range 0.1 -- 1 Myr (so that
$|\epsilon {\dot a}/{\dot \epsilon} a| \sim 1$).  The planetary IMF
must be nearly log-random, more specifically, close to the observed
planetary mass function with a moderate excess of lower mass planets
(they are the ones accreted or ejected). Finally, in order to not
overpopulate the (low $a$, high $\epsilon$) portion of parameter
space, close orbits must be circularized and hence the tidal quality
factor must lie in the estimated range $Q = 10^5 - 10^6$. If the
system parameters differ significantly from these values/ranges, then
this migration mechanism (in the form studied here) will not produce
the observed orbital elements of extrasolar planets.

Although this migration mechanism is promising, a number of issues
remain unresolved and should be considered in future work. One
important issue is the manner in which eccentricity is damped by the
circumstellar disk. The parametric treatment presented here indicates
that in order for the theory to produce results consistent with the
observed distributions of orbital elements, the ratio of the
eccentricity damping time scale to the migration (disk accretion) time
scale should lie in the range $\taued/\taudamp \approx 1/3 - 3$.
Numerical simulations of circumstellar disks interacting with planets
often provide a damping time scale near the low end of this range
(e.g., Nelson et al. 2000, Kley et al. 2004), whereas competing
analytic calculations suggest that eccentricity is not damped at all,
but rather is excited by the disk (Goldreich and Sari 2003, Ogilvie
and Lubow 2003). Since a particular disk cannot damp and excite
eccentricity at the same time, these two conflicting results define an
interesting problem for future study. The resolution of this issue
must allow for eccentricity damping time scales in the proper range
($\taued/\taudamp \approx 1/3 - 3$) if this migration scenario
represents the correct explanation for the observed extrasolar
planetary orbits.

The most fundamental challenge for the future is to determine which
migration scenario (or scenarios) acts to produce the observed orbits
of extrasolar planets. Any viable migration scenario must be able to
explain the observed distributions of orbital elements.  This paper
shows that Type II migration (modeled using parametric disk torques)
acting in combination with planet-planet scattering can produce these
observed distributions, but perhaps other mechanisms will also prove
to be successful -- there is not a uniqueness theorem for this
problem.  Discriminating among mechanisms might therefore rely on
secondary predictions.  Although the full consequences of the various
migration mechanisms have not been worked out, some preliminary
statements can be made: For example, if scattering processes play a
role in planet migration, then the scattering bodies will often be
ejected. More scattering leads to more ejected bodies.  The scenario
considered here (planet-planet scattering combined with Type II
migration through a circumstellar disk) implies that about one third
of the solar systems will eject a planet with (roughly) one Jovian
mass. For scenarios in which a circumstellar disk provides the
required eccentricity excitation but multiple planets are present,
planet-planet interactions tend to excite additional eccentricity and
planets are more readily ejected. As a specific example, we ran an
additional set of 375 simulations with the log-random IMF, $\taudamp$
= 0.3 Myr, and $\taued = -1$ Myr (corresponding to eccentricity
excitation); we found that 100\% of the systems ejected one of the
planets so that no multiple planet systems remained. For scenarios in
which migration is driven by scattering processes only, even more 
ejections are predicted. For example, in the case of ten planet
systems (AL2003), each solar system ejects 8 or 9 planets with
(roughly) Jovian characteristics. If the scattering bodies are
planetesimals (Murray et al. 1998), then each solar system would eject
several Jupiter masses worth of scattering bodies, but they would be
in rocky form. Thus, the currently discussed migration mechanisms
provide differing amounts of scattered debris. In the future,
additional signatures that discriminate between the various migration
scenarios should be identified and developed.
 
\bigskip   
\centerline{\bf Acknowledgments} 
\medskip 

We would like to thank Paul Butler, Gus Evrard, Debra Fischer, Greg
Laughlin, Man Hoi Lee, and Steve Lubow for beneficial discussions. We
also thank the referees for their many useful comments that improved
the paper.  This work was supported at the University of Michigan by
the Michigan Center for Theoretical Physics and by NASA through the
Terrestrial Planet Finder Mission (NNG04G190G) and the Astrophysics
Theory Program (NNG04GK56G0).

{} 

\newpage 
\bigskip
\centerline{\bf Table 1: Planetary Fate Probabilities} 
\medskip 

\begin{center}
\begin{tabular}{lcccc}
\hline 
\hline
$\taued$ & Ejection & Accretion & Collision & Survival\\
\hline 
Linear IMF\\
0.3 Myr & 0.526 & 0.332 & 0.031 & 0.112\\ 
1.0 Myr & 0.562 & 0.192 & 0.006 & 0.241\\ 
3.0 Myr & 0.548 & 0.163 & 0.003 & 0.286\\ 
\hline 
Log IMF\\
0.1 Myr  & 0.253 & 0.402 & 0.072 & 0.273\\
0.3 Myr  & 0.339 & 0.347 & 0.030 & 0.285\\ 
1.0 Myr  & 0.335 & 0.294 & 0.005 & 0.371\\ 
$\infty$ & 0.348 & 0.333 & 0.004 & 0.295\\ 
\hline 
Log IMF (2:1) \\ 
1.0 Myr & 0.288 & 0.349 & 0.011 & 0.352\\ 
\hline 
\hline 
\end{tabular}
\end{center} 

%\newpage 
\vskip 1.0truein 

\bigskip
\centerline{\bf Table 2: Planet Masses (in $m_J$)}  
\medskip 

\begin{center}
\begin{tabular}{lccccc}
\hline 
\hline
$\taued$ & Ejectors & Ejectees & Accreted & Accretion Surviver & 2-Planet Systems\\
\hline
Linear IMF\\
0.3 Myr & 3.67 $\pm$ 0.88 & 2.53 $\pm$ 1.36 & 1.01 $\pm$ 0.64 & 3.29 $\pm$ 1.08 & 1.27 $\pm$ 0.76 \\ 
1.0 Myr & 3.56 $\pm$ 0.96 & 2.21 $\pm$ 1.38 & 0.79 $\pm$ 0.65 & 3.25 $\pm$ 1.11 & 1.83 $\pm$ 1.11 \\ 
3.0 Myr & 3.47 $\pm$ 0.93 & 2.22 $\pm$ 1.51 & 0.65 $\pm$ 0.54 & 3.43 $\pm$ 1.06 & 1.76 $\pm$ 0.98 \\  
\hline
Log IMF\\ 
0.1 Myr  & 5.73 $\pm$ 1.89 & 1.60 $\pm$ 1.88 & 0.92 $\pm$ 1.04 & 3.25 $\pm$ 2.72 & 0.80 $\pm$ 0.82\\
0.3 Myr  & 5.19 $\pm$ 2.35 & 1.75 $\pm$ 2.06 & 0.60 $\pm$ 0.61 & 3.90 $\pm$ 2.75 & 0.69 $\pm$ 0.69 \\ 
1.0 Myr  & 4.92 $\pm$ 2.47 & 1.80 $\pm$ 2.08 & 0.51 $\pm$ 0.51 & 3.95 $\pm$ 2.70 & 0.78 $\pm$ 0.71 \\
$\infty$ & 5.03 $\pm$ 2.42 & 1.75 $\pm$ 1.93 & 0.45 $\pm$ 0.44 & 3.71 $\pm$ 2.84 & 0.87 $\pm$ 0.97 \\
\hline 
Log IMF (2:1)\\
1.0 Myr & 4.81 $\pm$ 2.16 & 1.62 $\pm$ 2.23 & 0.65 $\pm$ 0.91 & 3.78 $\pm$ 2.60 & 0.78 $\pm$ 0.74 \\ 
\hline 
\hline 
\end{tabular}
\end{center} 

\newpage  

\bigskip
\centerline{\bf Table 3: Semi-major Axes of Remaining Planets (in AU)}  
\medskip 

\begin{center}
\begin{tabular}{lcccc}
\hline 
\hline
$\taued$ & Ejectors & Accretion Surviver & 2-Planet Systems & All Survivors\\
\hline
Linear IMF\\
0.3 Myr & 2.26 $\pm$ 1.75 & 0.18 $\pm$ 0.22 & 0.42 $\pm$ 0.41 & 1.22 $\pm$ 1.24 \\ 
1.0 Myr & 2.50 $\pm$ 2.69 & 0.70 $\pm$ 1.58 & 0.37 $\pm$ 0.34 & 1.37 $\pm$ 0.53 \\ 
3.0 Myr & 2.29 $\pm$ 1.11 & 1.00 $\pm$ 1.10 & 0.60 $\pm$ 0.44 & 1.37 $\pm$ 0.87 \\  
\hline
Log IMF\\ 
0.1 Myr  & 3.10 $\pm$ 6.18 & 0.21 $\pm$ 1.36 & 0.45 $\pm$ 0.44 & 0.97 $\pm$ 3.05 \\
0.3 Myr  & 2.48 $\pm$ 0.94 & 0.23 $\pm$ 0.36 & 0.41 $\pm$ 0.44 & 0.90 $\pm$ 0.60 \\ 
1.0 Myr  & 2.62 $\pm$ 2.68 & 0.62 $\pm$ 0.60 & 0.53 $\pm$ 0.44 & 1.05 $\pm$ 1.38 \\
$\infty$ & 2.68 $\pm$ 1.89 & 1.46 $\pm$ 0.91 & 1.26 $\pm$ 0.68 & 1.69 $\pm$ 1.18 \\
\hline 
Log IMF (2:1)\\
1.0 Myr & 2.61 $\pm$ 0.94 & 0.53 $\pm$ 0.88 & 0.49 $\pm$ 0.39 & 0.95 $\pm$ 0.69 \\ 
\hline 
\hline 
\end{tabular}
\end{center} 

%\newpage 
\vskip1.0truein 

\bigskip
\centerline{\bf Table 4: Eccentricities of Remaining Planets}  
\medskip  
\begin{center}
\begin{tabular}{lcccc}
\hline 
\hline
$\taued$ & Ejectors & Accretion Surviver & 2-Planet Systems & All Survivors\\
\hline
Linear IMF\\
0.3 Myr & 0.46 $\pm$ 0.28 & 0.28 $\pm$ 0.22 & 0.46 $\pm$ 0.25 & 0.41 $\pm$ 0.26 \\ 
1.0 Myr & 0.43 $\pm$ 0.26 & 0.45 $\pm$ 0.29 & 0.68 $\pm$ 0.19 & 0.53 $\pm$ 0.24 \\ 
3.0 Myr & 0.40 $\pm$ 0.27 & 0.43 $\pm$ 0.32 & 0.79 $\pm$ 0.14 & 0.58 $\pm$ 0.23 \\  
\hline
Log IMF\\ 
0.1 Myr & 0.24 $\pm$ 0.25  & 0.041 $\pm$ 0.070 & 0.30 $\pm$ 0.22 & 0.20 $\pm$ 0.19 \\
0.3 Myr & 0.25 $\pm$ 0.23  & 0.14 $\pm$ 0.18 & 0.44 $\pm$ 0.25 & 0.31 $\pm$ 0.23 \\ 
1.0 Myr  & 0.28 $\pm$ 0.26 & 0.34 $\pm$ 0.26 & 0.65 $\pm$ 0.22 & 0.49 $\pm$ 0.24 \\
$\infty$ & 0.29 $\pm$ 0.25 & 0.35 $\pm$ 0.35 & 0.83 $\pm$ 0.13 & 0.55 $\pm$ 0.24 \\
\hline 
Log IMF (2:1)\\
1.0 Myr & 0.26 $\pm$ 0.26 & 0.38 $\pm$ 0.23 & 0.65 $\pm$ 0.21 & 0.46 $\pm$ 0.23 \\ 
\hline 
\hline 
\end{tabular}
\end{center}

\newpage 
\bigskip
\centerline{\bf Table 5: Percent of Total Time Spent in 2:1/3:1 Resonances} 
\medskip 

\begin{center}
\begin{tabular}{lccccc}
\hline 
\hline
$\tau_{\rm damp}$ & TimeStop & Eject & Accrete & Total & Total(alt) \\
\hline 
Linear IMF\\
0.3 Myr  & 47/50 &  2/83 & 18/69 & 22/67 & 12/73\\
1.0 Myr  & 25/72 &  9/69 & 20/68 & 20/70 & 13/70\\
3.0 Myr  & 21/77 & 11/68 & 19/67 & 19/73 & 13/71\\
\hline 
Log IMF\\
0.1 Myr  & 64/32 & 10/61 & 40/54 & 49/44 & 34/52\\
0.3 Myr  & 70/25 &  8/63 & 33/46 & 51/35 & 33/46\\ 
1.0 Myr  & 70/26 & 17/55 & 34/46 & 55/34 & 37/44\\ 
$\infty$ & 62/33 & 15/56 & 43/37 & 51/37 & 32/46\\ 
\hline 
Log IMF (2:1)\\
1.0 Myr  & 100/0 & 94/0.3 & 100/0 & 100/0 & 94/0.1\\
\hline 
\end{tabular}
\end{center} 

%\newpage 
\vskip 1.0truein 

\bigskip
\centerline{\bf Table 6: Linear Correlation Coefficient between 
$\epsilon$ and $i$} 
\medskip  
\begin{center}
\begin{tabular}{lcccc}
\hline 
\hline
$\tau_{\rm damp}$ & Survival & Ejection & Accretion & Total\\
\hline 
Linear IMF\\
0.3 Myr & 0.62  (44) & 0.53  (99) & 0.61  (96) & 0.50 (214)\\ 
1.0 Myr & 0.064 (260) & 0.43 (294) & 0.57 (103) & 0.31 (660)\\ 
3.0 Myr & 0.061 (190) & 0.39 (180) & 0.76  (54) & 0.27 (425)\\ 
\hline 
Log IMF\\
0.1 Myr  & 0.69 (428) & 0.59 (221) & 0.55 (333) & 0.64 (1034)\\
0.3 Myr  & 0.58 (456) & 0.53 (258) & 0.57 (274) & 0.54 (1013)\\ 
1.0 Myr  & 0.30 (566) & 0.56 (248) & 0.33 (224) & 0.45 (1038)\\ 
$\infty$ & 0.28 (398) & 0.40 (228) & 0.66 (238) & 0.46 (867)\\ 
\hline 
Log IMF (2:1)\\
1.0 Myr  & 0.36 (198) & 0.56 (80) & 0.60 (98) & 0.51 (379)\\
\hline 
\hline 
\end{tabular}
\end{center} 

\newpage 

\newpage 

\newpage

\newpage

\newpage 

\newpage 

\newpage 

\newpage 

\newpage 

\newpage 

\newpage 

\newpage 

\newpage 

\newpage 

\newpage 

\end{document}